\documentclass[prl,aps,twocolumn,eqsecnum,a4paper,ltxgrid]{revtex4}

\usepackage{graphicx}
\usepackage[sumlimits]{amsmath}
\usepackage{amsfonts}
\usepackage{amssymb}
\usepackage{dcolumn}
\usepackage{subdepth}
\usepackage{color}
\usepackage{enumitem}

\newcommand{\VB}{V_{\mathrm{B}}}
\newcommand{\VS}{V_{\mathrm{S}}}
\newcommand{\VSt}{V'_{\mathrm{S}}}
\newcommand{\VSB}{V_{\mathrm{S-B}}}

\newcommand{\Eq}[1]{Eq.~(\ref{#1})}

\newcommand{\Fig}[1]{Fig.~\ref{#1}}
\newcommand{\Sec}[1]{Sec.~\ref{#1}}

\newcommand{\xid}{\xi_\mathrm{CL}}
\newcommand{\xilp}{\xi_\mathrm{LP}}
\newcommand{\xirs}{\xi_\mathrm{RS}}
\newcommand{\xirb}{\xi_\mathrm{RB}}
\newcommand{\xidh}{\hat{\xi}_\mathrm{CL}}
\newcommand{\xilph}{\hat{\xi}_\mathrm{LP}}
\newcommand{\xirsh}{\hat{\xi}_\mathrm{RS}}
\newcommand{\xirbh}{\hat{\xi}_\mathrm{RB}}
\newcommand{\xilpt}{\tilde{\xi}_\mathrm{LP}}
\newcommand{\xirst}{\tilde{\xi}_\mathrm{RS}}
\newcommand{\xirbt}{\tilde{\xi}_\mathrm{RB}}
\newcommand{\cm}{cm$^{-1}$}
\newcommand{\diff}{\mathrm{d}}
\renewcommand{\i}{\mathrm{i}}
\begin{document}

\title{Parametrizing linear generalized Langevin dynamics from explicit molecular dynamics simulations}

\author{Fabian Gottwald}
\author{Sven Karsten}
\author{Sergei D. Ivanov}
\email{sergei.ivanov@uni-rostock.de}
\author{Oliver K\"uhn}
\address{Institute of Physics, Rostock University, Universit\"atsplatz 3, 18055 Rostock, Germany}

\begin{abstract}

%SI2: Introductory sentence added and everything reworked according to 
Fundamental understanding of complex dynamics in many-particle systems on the atomistic
level is of utmost importance.
Often the systems of interest are of macroscopic size but can be partitioned into
few important degrees of freedom which are treated most accurately
and others which constitute a thermal bath.
Particular attention in this respect attracts the linear generalized Langevin equation (GLE),
which can be rigorously derived by means of a linear projection (LP) technique.
Within this framework a complicated interaction with the bath can be reduced to a single memory kernel.
This memory kernel in turn is parametrized for a particular system studied,
usually by means of time-domain methods based on explicit  molecular dynamics data.
Here we discuss that this task is most naturally achieved in frequency domain
and develop a Fourier-based parametrization method that outperforms its time-domain analogues.
Very surprisingly, the widely used rigid bond method turns out to be inappropriate in general.
Importantly, we show that the rigid bond approach leads to a systematic underestimation of
relaxation times, unless the system under study consists of a harmonic bath bi-linearly coupled to the
relevant degrees of freedom.

\end{abstract}
\date{\today}
\maketitle

%\twocolumn
\section{Introduction}

%FG: First section of thesis intro. Question whether we want to put so much emphasis on spectroscopy
%FG: If yes we need to put some milestone papers about (experimental) spectroscopy
Studying complex dynamics of many-particle systems has become
one of the main goals in modern molecular physics. The fundamental
understanding of the underlying 
%SI2:
microscopical
processes requires the interplay of elaborate
experimental techniques and sophisticated theoretical approaches.
Experimentally, (non-)linear spectroscopy revealed itself as a powerful tool for probing
the dynamics and for determining the characteristic timescales,
%SI2: To give examples
such as dephasing/relaxation times and reaction rates to name but two.
%SI2:
%for the processes of interest.
%
For interpreting the experimental spectra theoretical models are needed
which can give insight into the atomistic dynamics.
% on a microscopic level.
%
Often, a reduction of the description 
%SI: Oliver, 'a few' means 'some'. Please don't correct back.
to few variables is convenient in many cases since this can not only ease the interpretation, but enable the identification of key properties~\cite{Kuehn2011}.
Such a reduced description can formally be obtained from the so-called system-bath partitioning,  where only a small subset of degrees of freedom (DOFs), referred to as system, is considered as important for describing a physical process under study. All the other DOFs, referred to as bath, are regarded as irrelevant in the sense that they might influence the time evolution of the system but do not 
\textit{explicitly} enter any dynamical variable of interest. 
%SI2: Also to specify what we are talking about.
Practically, such a separation is often natural, for instance, when studying a reaction with a clearly defined
reaction center or solute dynamics in a solvent environment.
Further, reduced equations of motion (EOMs) for the system DOFs can be derived in which the influence of the bath
is limited to dissipation and fluctuations.

The most simple formulation of this idea is provided by the Markovian Langevin equation,
where dissipation and fluctuations take the form of static friction and stochastic white noise, respectively~\cite{Lemons1997,Ornstein1930,ZwanzigBook2001}. 
Situations where memory effects become important are accounted for by the generalized Langevin equation (GLE)~\cite{Mori1965,Zwanzig1973,Kawasaki1973} via a frequency-dependent friction and a stochastic force with a finite correlation time. 
This generalized equation has been
%SI2:
employed
 %involved,
  for instance, in the theory of vibrational relaxation for estimating characteristic relaxation times~\cite{Whitnell1990,Benjamin1993,Tuckerman1993a,Gnanakaran1996a}, reaction rates~\cite{Grote-JCP-1980}
 and for thermostatting purposes~\cite{Ceriotti2009,Ceriotti2009a, Ceriotti2010};
 see e.g.\ Refs.~\cite{Abe-PR-1996,Tanimura2006} for review.

The microscopic origin of the GLE can be rationalized starting from different standpoints. 
First, it can be rigorously derived from the so-called Caldeira-Leggett (CL) model, where the environment is assumed to be a collection of independent harmonic oscillators bi-linearly coupled to the system~\cite{Zwanzig1973,Caldeira-PRL-1981,Caldeira-AP-1983, ZwanzigBook2001,Mukamel1995}.
This model has been widely used in analyzing and interpreting (non-)linear spectroscopic experiments on systems in condensed phase,
termed multi-mode Brownian oscillator (MBO) model in this context~\cite{Palese1996,Okumura1997,Woutersen1999,Toutounji2002,Tanimura2009,Joutsuka2011}.
The second, more formal ansatz is to employ projection operator techniques in order to recast the system's EOMs into linear or non-linear GLE forms~\cite{Mori1965,Kawasaki1973,Kawai2011,ZwanzigBook2001}.
In the former, the resulting system potential is effectively harmonic, whereas in the latter the system potential is formed by a (non-linear) mean-field potential.
In this approach noise and dissipation can be mathematically defined as (non-)linearly projected quantities.

In any case, practical use of the GLE can only be made in connection with a stochastic model for the noise term
being the main assumption of the formalism.
The general advantage of the stochastic GLE is that dissipation and the statistical properties of the noise are entirely described 
by the so-called memory kernel being simply a function of time.
If such a memory kernel can be obtained for a real system then  the full quantum-mechanical treatment of the bath can be performed analytically,
leading to a quantum version of the GLE~\cite{Cortes1985a,Ford-JSP-1987,Ford-PRA-1988, McDowell2000, Banerjee2004}.
Alternatively, a density matrix theory either via the Feynman-Vernon influence functional approach~\cite{Feynman-AP-1963,Caldeira-PhysicaA-1983} or hierarchy type EOMs~\cite{Dijkstra-PRL-2010,Suess:2014gz} can be employed.
Further, numerical methods that solve the Schr\"odinger equation in many dimensions~\cite{Kuehn-PRP-2006} or
various quantum-classical hybrid schemes~\cite{egorov99_5238} can be used.
Finally, the machinery for a purely classical treatment of the GLE is provided by the method of colored noise thermostats~\cite{Ceriotti2009a,Ceriotti2010} or similar techniques~\cite{Baczewski2013, Stella2014}.
%FG: Ref to review
Therefore the GLE formalism has become a popular tool for assigning system properties in macroscopic environments.

However, establishing a connection between a real molecular system and a GLE might not be straightforward.
In a recent study we have shown that in condensed phase the mapping between the two
can be established only for the effectively harmonic GLE derived by means of linear projection (LP)
operator techniques~\cite{Ivanov-arxiv-2014}.
All other possible mappings onto the GLE where the system force is kept anharmonic turned out inapplicable
due to the so-called invertibility problem.
Hence, in this work we limit ourselves exclusively to parametrizing the linear GLE.

Common approaches to calculating memory kernels involve classical molecular dynamics (MD) simulations where the environment is explicitly taken into account. A very popular scheme is to obtain the memory kernel from the time-correlation function (TCF) of the forces exerted on a frozen system coordinate,  referred to as the rigid bond approach~\cite{Whitnell1990,Benjamin1993,Gnanakaran1996a,Kuehn-PCCP-2003}.
However, Berne et al.~\cite{Berne1990} have shown that this ansatz is only correct when the system frequency is much larger than the bath ones. If such a frequency separation is not given, one can determine the memory kernel from a Volterra integro-differential equation for the momentum-autocorrelation function (MAF). 
Practically, the memory kernel can be computed from explicit MD MAFs involving discretization schemes in time domain~\cite{Berne-ACP-1970,Berkowitz1981,Lange2006,Shin2010}. An example of this ansatz is the method introduced by Berne and Harp who have employed polynomial interpolations of the MAF in order to calculate necessary derivatives~\cite{Berne-ACP-1970,Straub1987,Berne1990,Tuckerman1993a}. 

Another idea is based on Laplace domain techniques~\cite{Goodyear1996,Lange2006,Baczewski2013}. Here, the Volterra integro-differential equation is transformed into Laplace domain, resulting in an algebraic expression for the Laplace-transformed memory kernel.
However, the practical application of this method has not been discussed in detail. In contrast, this technique has 
been criticized due to the significant difficulties of the numerical Laplace back transform~\cite{Shin2010}.

In this paper we demonstrate that it is more convenient to parametrize the GLE in frequency domain,
where a memory kernel turns into a spectral density.
We present a new Fourier-based method, 
which provides a direct and robust way for calculating spectral densities of realistic solute dynamics in liquid solvents
and avoids the aforementioned numerical problems.
It is shown that existing time-domain techniques suffer from either numerical or conceptual deficiencies.
Very surprisingly, the well-established rigid bond approach turns out to be inapplicable, even when the
frequency separation is provided.

The paper is organized as follows.
After this introduction,  the GLE formalism is described in detail.
Then the rigid bond approach and the method by Berne and Harp,~\cite{Berne-ACP-1970}
which are widely used for calculating the memory kernel
from explicit MD simulation data are reviewed.
Further, we present in detail a new efficient scheme that enables calculating the memory kernel in Fourier space.
Finally, we demonstrate that our procedure gives reasonable spectral densities for realistic solute dynamics
on the example of two hydrogen bonded systems: HOD in H$_2$O, 
 and the ionic liquid $[\mathrm{C}_{2}\mbox{mim}][\mbox{NTf}_{2}]$.
%
%SI2: Phrase is made more clear.
The model system introduced by Berne et al.~\cite{Berne1990}, which consists of a diatomic in 
an atomic gas, termed A$_2$ in A, is considered for cross-checking.
A comparison of our procedure against the rigid system approach and the method of Berne and Harp is provided,
and the failure of the rigid bond approach is analyzed in detail 
followed by conclusions and outlook.

\section{Generalized Langevin equations}

In the following we consider a classical one-dimensional system with a mass $m$, coordinate $x$ and conjugate momentum $p$,
which undergoes Brownian motion in a classical bath described by coordinates $\{ Q_i \}$ and momenta $\{ P_i \}$.
The total open system is without any loss of generality assumed to be partitioned into the system, $\VS(x)$, and the
bath, $\VB(\{ Q_i \})$, coupled  via a system-bath interaction, $\VSB(x, \{ Q_i \})$.

\subsection{The linear GLE}

A mathematically rigorous approach for deriving reduced EOMs is to employ LP operators,
which project a dynamical variable $A(x,p;\{ Q_i , P_i \})$ depending on the full set of system and bath coordinates onto the linear subspace spanned by $x$ and $p$.~\cite{Mori1965,Kawasaki1973,ZwanzigBook2001}
The resulting linear GLE reads
\begin{eqnarray}
\label{eq:GLE}
\dot{p}(t) & = & -\frac{kT}{\langle x^2 \rangle}x(t) -\intop_{0}^{t}\xilp(t-\tau)p(\tau)\diff\tau+R(t)\nonumber \\
\dot{x}(t) & = & \frac{p(t)}{m}
\enspace,
\end{eqnarray}
where $T$ stands for the temperature, $k$ is the Boltzmann constant and $\langle ... \rangle$ denotes canonical averaging.
The dissipative force is thus given by a convolution of the momentum with a memory kernel, $\xilp(t)$,
which is a real function decaying on a finite timescale.
The noise term, $R(t)$, formally contains an explicit dependence on all bath DOFs
and is related to the memory kernel via the fluctuation-dissipation theorem (FDT)
\begin{equation}
\label{eq:FDT}
\xilp(t) = \frac{\left\langle R(0)\exp[\i L_\mathrm{LP}t] R(0)\right\rangle}{mkT}
\enspace ,
\end{equation}
where $L_\mathrm{LP}$ is the Liouville operator that describes the linearly projected time evolution;
note that it does not correspond to the real Hamiltonian flow.
In practical applications one sacrifices the deterministic time evolution of the noise and mimics this term by a stochastic zero-centered Gaussian process $R(t)$ keeping the FDT as its main statistical property~\cite{Berkowitz1983,Ceriotti2010,Baczewski2013}.
The GLE in \Eq{eq:GLE} then becomes a non-Markovian stochastic differential equation, where memory effects in the noise are incorporated via the FDT, \Eq{eq:FDT}.
The Markovian limit $\xilp(t) \rightarrow \xi_0 \delta(t)$ can be obtained if the decay of the memory kernel is much faster than a characteristic timescale, e.g.~a vibrational period, of the system.
The great advantage of the stochastic GLE is that the implicit characterization of the bath is simply given by the memory kernel $\xilp(t)$.
For an easier interpretation it is convenient to consider $\xilp(t)$ in frequency domain, where it is referred to as the spectral density
\begin{equation}
\label{eq:SD}
\xilph(\omega)=\intop_0^{\infty} \exp[-i\omega t]\xilp(t)\diff t
\enspace .
\end{equation}
Here and in the following hat denotes the half-sided Fourier transform.
It should be stressed at this point that the term which is linear in $x$ in \Eq{eq:GLE}, usually referred to as the
 mean intramolecular force of the system,~\cite{Tuckerman1993a,Gnanakaran1996a}
is a consequence of the LP.
This implies that one can interpret the projected system part as an effective harmonic oscillator with the frequency 
\begin{equation}
\label{eq:frequency}
\bar{\omega}^2 \equiv \frac{kT}{m\langle x^2 \rangle}
\end{equation}
even though the original system potential can be arbitrarily anharmonic.
The anharmonicity is formally projected onto the bath and, hence, incorporated into the noise term and the memory kernel. 
Thus, using this model to disentangle, e.g.\ anharmonicity and mode coupling,
by means of non-linear spectroscopies is doomed to fail from the outset.
Still, if one is interested only in linear properties, like transport coefficients or linear absorption spectra,
the GLE gives the correct description~\cite{Ivanov-arxiv-2014} and provides a very simple and intuitive model.
In the general case the linear GLE does not reflect all dynamical features of the real system and, thus, one partly looses the true atomistic picture when using it.

\subsection{The Caldeira-Leggett Model}
Another way to derive a GLE is to employ the CL model that has enjoyed popularity in condensed phase spectroscopy~\cite{Palese1996,Okumura1997,Woutersen1999,Toutounji2002,Tanimura2009,Joutsuka2011}.
This model assumes that the bath consists of independent harmonic oscillators
bi-linearly coupled to the system.
The total potential energy for the model reads
\begin{equation}
\label{eq:CL Hamiltonian}
V(x,\{Q_i\})=\VS(x)+\sum_{i}\frac{1}{2}\omega_{i}^{2}Q^2_{i}-\sum_{i}g_{i} Q_i x 
\end{equation}
with the bath frequencies $\omega_{i}$, bath masses set to unity and the coupling strengths
$g_{i}$.~\cite{Caldeira-PRL-1981,Caldeira-AP-1983,grabert88_115}
Often the square on the right hand side of \Eq{eq:CL Hamiltonian} is completed
%
%\begin{equation}
%\label{eq:MBO Hamiltonian}
%\begin{split}
%\VS(x)+&\VB(\{Q_i\})+\VSB(x,\{Q_i\})\equiv \\
%&\VSt(x) + \sum_{i}\frac{1}{2}\omega_{i}^{2}\left(Q_{i}-\frac{g_{i}}{\omega_{i}^{2}}x\right)^{2}
%\end{split}
%\end{equation}
%
causing thereby a harmonic correction to the system potential $\VS(x)$
\begin{equation}
\label{eq: corrected potential}
\VSt(x) \equiv \VS(x) - \frac{1}{2} \sum_i \frac{g_i^2}{ \omega_i^2} x^2 \enspace .
\end{equation}
%
%SI: Oliver, we successfully confused ourselves about what is 'consistent' and what is not. So let's not discuss it here,
%SI: but we should discuss it among us!
%This version of the CL model is often referred to as the multi-mode Brownian oscillator (MBO) model in spectroscopy community~\cite{Mukamel1995} and it differs from the original CL model only by the partitioning of the system and the bath.
%SI: Commented out
%Note that the MBO model in spectroscopy is commonly used in the context of an adiabatic separation of high-frequency vibrations (or electronic transitions) from the vibrational modes described by the MBO model~\cite{stenger01_027401}. The model includes the well-known cases of Kubo line shape theory as slow- and fast modulation limits~\cite{kubo69_101}.
%
The corresponding GLE implied by the CL model can be derived  both in classical~\cite{Zwanzig1973,ZwanzigBook2001} and 
quantum~\cite{Ford-JSP-1987,Ford-PRA-1988} domains.
Limiting ourselves to the classical description, the derivation can be straightforwardly performed without any further approximations
by integrating the EOMs for the bath, yielding
\begin{equation}
\label{eq:CL-GLE}
\dot{p}(t) = -\frac{\partial \VSt}{\partial x}-\intop_{0}^{t}\xid(t-\tau)p(\tau)\diff \tau+R(t)
\enspace ;
\end{equation}
%
%SI2: Can be moved out.
%note that $\xid(t)$ was referred to as $\xi_\mathrm{D}(t)$ in Ref.~\cite{Ivanov-arxiv-2014}.
%
Within the CL model the real part of the spectral density, \Eq{eq:SD}, can be interpreted as the bath modes'
 density of states weighted with the strength of the coupling to the system
\begin{equation}
\label{eq:CL-SD}
\Re \xidh(\omega)\equiv \sum_{i}\frac{g_{i}^{2}}{\omega_{i}^{2}}\delta(\omega-\omega_{i}) \enspace .
\end{equation}
By comparing \Eq{eq:CL-SD} and \Eq{eq: corrected potential} one finds that the frequency renormalization term is given by 
%
%SI: This is by the way $\xi(0)$ that Berne and Tuckerman are looking at.
\begin{equation}
\VSt(x)=\VS(x) - \frac{x^2}{2} \int\limits_0^\infty  \Re \xidh(\omega)\diff \omega\,  \enspace .
\end{equation}
This frequency renormalization term thus induces a redshift of the system frequency due to interactions with the harmonic oscillator bath. 
Comparing \Eq{eq:GLE} to \Eq{eq:CL-GLE} immediately suggests that the only possibility for them to coincide
is to set the model system potential $\VSt (x) =   kT x^2/(2\langle x^2 \rangle) $.

\section{Parametrizing a Spectral Density}
 
\subsection{Formulation of the problem}
In order to utilize the framework of a stochastic GLE one has to find a memory kernel
that correctly mimics the features of the bath.
We start the consideration with the linear GLE, \Eq{eq:GLE}.
Multiplying it with $p(0)$, taking the canonical ensemble average and noting that the correlation function with the noise term vanishes by construction, yields a Volterra integro-differential equation for the 
normalized MAF defined as $C_{pp}(t) \equiv \langle p(0)p(t) \rangle /\langle p^2 \rangle$
\begin{equation}
\label{eq:Volterraequation-Cpx}
\dot{C}_{pp}(t)=-m\bar{\omega}^2C_{px}(t) - \intop_0^t \xilp(t-\tau)C_{pp}(\tau) \diff \tau \enspace,
\end{equation}
where the function $C_{px}(t)\equiv\langle p(0)x(t) \rangle /\langle p^2 \rangle$ denotes the momentum-position cross correlation.
The aforementioned equation can be simplified by using the relation 
\begin{equation}
\label{eq:Cpx-Cpp}
C_{px}(t)=\frac{1}{m} \intop_0^t C_{pp}(\tau)\diff\tau
\enspace ,
\end{equation}
which yields
\begin{equation}
\label{eq:Volterraequation}
\dot{C}_{pp}(t)=-\intop_0^tK_\mathrm{LP}(t-\tau)C_{pp}(\tau) \diff \tau \enspace,
\end{equation}
with the shifted memory kernel
\begin{equation}
\label{eq:K_t}
K_\mathrm{LP}(t) \equiv \xilp(t)+\bar{\omega}^2
\enspace .
\end{equation}
The basic idea is to invert this equation to obtain the memory kernel $K_\mathrm{LP}(t)$ and, hence,
the spectral density $\hat{K}_{\mathrm{LP}}(\omega)$.
As an input, the MAF calculated from an explicit MD simulation is needed.

%SI: Now CL-GLE
Performing the aforementioned procedure with \Eq{eq:CL-GLE} as a starting point, yields another Volterra equation
\begin{equation}
\label{eq:Volterraequation-CL}
\dot{C}_{pp}(t)= C_{pF}(t) - \intop_0^t\xid(t-\tau)C_{pp}(\tau) \diff \tau
\enspace,
\end{equation}
where $C_{pF}(t)$ is momentum-system force correlation function.
Thus these two functions, $C_{pp}(t)$ and $C_{pF}(t)$, can be used to parametrize $\xid(t)$.
It should be stressed that a self-consistent procedure can be established only if the system is of the CL form.
In general, there exist infinitely many pairs of the TCFs that correspond to the given spectral density,
which is in the heart of the invertibility problem.~\cite{Ivanov-arxiv-2014}
Still, the obtained spectral density is unique and can be used for comparison, see \Sec{sec:A2inA}.

\subsection{Berne and Harp Method}

The method presented in this section approaches the iterative solution for  $K_\mathrm{LP}(t)$
from \Eq{eq:Volterraequation} in the time domain
has been introduced by Berne and Harp in 1970~\cite{Berne-ACP-1970};
due to stability reasons, the starting point here is the derivative of \Eq{eq:Volterraequation}
rather than the integro-differential \Eq{eq:Volterraequation}  itself.
Following Ref.~\cite{Berne-ACP-1970}, the method consists of three steps.
First, the corresponding finite difference scheme is formulated
by making a variable substitution $\tau \to \tau -t$ in \Eq{eq:Volterraequation}
in order to shift the $t$-dependence from the kernel, 
followed by a differentiation with respect to $t$ resulting in
\begin{eqnarray}
\label{eq:diff_Volterraequation}
K_\mathrm{LP}(m \Delta t) = -\ddot{C}_{pp}(m \Delta t) &  \nonumber\\
-  \Delta t \sum_{j=0}^{m} w_j  K_\mathrm{LP}(j \Delta t)\!\!&\!\!\dot{C}_{pp}\left ( (m-j) \Delta t \right )\enspace .
\end{eqnarray}
Here the time integration $ \int_{0}^{t} \diff \tau K_\mathrm{LP}(\tau) \dot{C}_{pp}(t-\tau)$ is approximated with the help of the Gregory formula
employing the weights $ w_j $ and MD timestep $\Delta t$. 
Note that the Gregory formula is advantageous to the Simpson rule, since it does not rely on the parity of the number of integrated points~\cite{Todd1962}.

Second, the MAF is interpolated by a $2n$th order polynomial  $ \bar{C}_k(t)=\sum_{j=0}^{2n} a^{(k)}_j t^j $
in the vicinity of $k$-th time step such that  
$\bar{C}_k((k+j)\Delta t)={C}_{pp}((k+j)\Delta t)$,
for $j=-n,-n+1,\ldots, n-1,n$.
The derivatives that enter \Eq{eq:diff_Volterraequation} are then approximated by
the derivatives of the polynomial $\bar{C}_k$ at these time points.

Third, one determines the initial values of $K_\mathrm{LP}$ at first four time instances.
The first value $K_\mathrm{LP}(0)=\langle\dot{p}^2(0)/p^2(0)\rangle$ is directly calculated from averaged MD data.
Subsequently, the remaining starting values $ K_\mathrm{LP}(\Delta t), K_\mathrm{LP}(2 \Delta t), K_\mathrm{LP}(3 \Delta t) $
are accurately determined by applying the so-called Day's method to \Eq{eq:diff_Volterraequation}.~\cite{Day1967}

\subsection{Rigid Bond Approach}
\label{sec:Rigid bond}

Another time domain technique which has been widely used in the literature, e.g.~in studies of vibrational relaxation~\cite{Whitnell1990,Berne1990,Gnanakaran1996a}, is based on approximating $\xilp(t)$ by $\xirb(t)$
calculated as 
\begin{equation}
\label{eq:rigid bond}
\xirb(t)=\frac{\langle R(0)\exp[ \i L_\mathrm{RB} t] R(0) \rangle}{mkT} 
\enspace .
\end{equation}
Here, $ L_\mathrm{RB}$ is the Liouvillian corresponding to the system with the frozen bond and the noise is
calculated via explicit MD simulations.
%
%This avoids contributions from the dissipative term (which are proportional to the momentum $p(t)$).
%
If a system coordinate $x$ is chosen as a single bondlength, as in the practical applications considered here,
then the noise $R(t)$ is obtained as
\begin{equation}
R(t)=\mu \cdot \vec{n}_{12}\cdot \left [ \frac{\vec{F}_1(t)}{m_1} - \frac{\vec{F}_2(t)}{m_2} \right ]
\enspace ,
\end{equation}
where $\vec{F}_1(t), \vec{F}_2(t)$ are the forces exerted on the two bonded atoms by their surrounding,
 $m_1, m_2$ are their masses and $\mu$ is the reduced mass of the atom pair.
The vector $\vec{n}_{12}$ is the unit bond vector pointing from atom 1 to atom 2.

In general, fixing the system's coordinate has an influence on the energy flow
between system and bath thereby affecting the memory kernel.
The consequences have been extensively discussed by Berne et.\ al~\cite{Berne1990}
and are briefly summarized below.
In general, there are two approximations behind the rigid bond approach, 
which can be formulated in terms of  the memory kernels $\xilp(t)$, $\xirb(t)$
defined in Eqs.~(\ref{eq:FDT}, \ref{eq:rigid bond}) and the time correlation function
due to the Hamiltonian flow of the real system
\begin{equation}
\label{eq:mem_kernels}
\xirs(t)= \frac{\langle R(0)\exp{(\i L_\mathrm{RS} t)}R(0)\rangle}{mkT} 
\enspace  .
\end{equation}
As it was shown by the authors, a simple relation between $\xirs(t)$ and $\xilp(t)$ can be obtained
in Laplace domain
\begin{equation}
\label{eq:RS-LP}
\xirst(s)=\frac{\xilpt(s)}{1+s/(\bar{\omega}^2+s^2)\xilpt(s)}
\enspace ,
\end{equation}
where tilde denotes the Laplace transform throughout the manuscript.
Taking the limit $\bar{\omega}\to\infty$ in \Eq{eq:RS-LP} naturally results in 
$\lim_{\bar{\omega}\to \infty}\xirst(s)=\lim_{\bar{\omega}\to \infty}\xilpt(s)$.
As a matter of fact, the limit also directly corresponds to the rigid bond dynamics,
$\xirbt(s)=\lim_{\bar{\omega}\to \infty}\xirst(s)$.
This can be most easily understood in the
time domain, namely, if the memory kernel varies slowly on the timescale of the fast oscillations of the system coordinate,
then the dissipative term in the GLE would vanish, as it is required by the rigid bond approach.
Combining the two 
%SI2:
expressions obtained in the limit $\bar{\omega}\to\infty$
%limits
yields the aforementioned two approximations $\xilp(t) \approx \xirs(t) \approx \xirb(t)$,
see \Sec{sec:timedomain} for further discussion.

Importantly, if the system studied would indeed be of the CL form,
then the match $\xid(t)= \xirb(t)$ becomes exact
and thus one can exclude $\xirs(t)$ from consideration.
The memory kernels $\xid(t)$ and $\xilp(t)$ coincide exclusively for harmonic system potential, $\VSt(x)$,
see \Sec{sec:A2inA}.

\subsection{Motivation for a Fourier Method}
\label{sec:Motivation}
The aforementioned methods are exclusively time domain techniques.
Another possibility that has been suggested in the literature is based on a transform of \Eq{eq:Volterraequation}
into Laplace domain.~\cite{Goodyear1996,Lange2006,Baczewski2013} 
This ansatz results in an algebraic equation
\begin{equation}
\label{eq:Volterra in Laplace domain}
s \tilde{C}_{pp}(s)-C_{pp}(t=0)=-\tilde{K}_\mathrm{LP}(s)\tilde{C}_{pp}(s)
\enspace .
\end{equation}
Note that a differentiation in time domain results in a multiplication by $s$ in Laplace domain
and the convolution in \Eq{eq:Volterraequation} thereby turns into a simple product.
Despite its apparent simplicity, the detailed implementation of the method has not been carried out to our best knowledge,
owing to numerical instabilities of a Laplace back transform.~\cite{Shin2010}.
Further we show that both time and Laplace domain methods are not natural for the present purpose.

\begin{figure}[t]
\includegraphics[width=0.99\columnwidth]{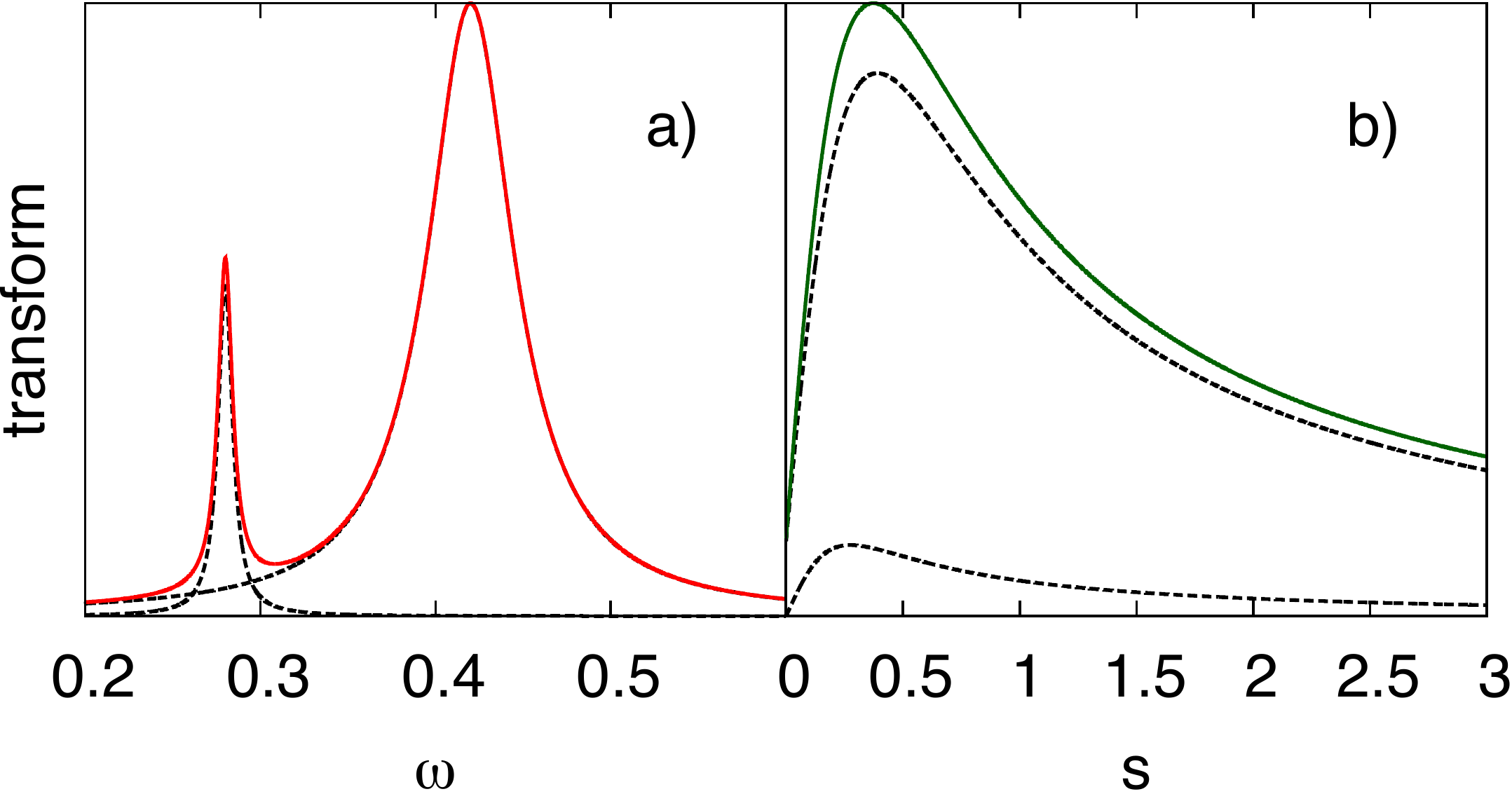}
\caption{\label{fig:laplace-vs-fourier}
Fourier transform (left panel) and Laplace transform (right panel) of a memory kernel consisting of a superposition of two exponentially damped cosine functions. Their individual Laplace/Fourier counterparts are plotted in black dotted lines.}
\end{figure}

First, numerical algorithms for GLE simulations often require to fit the memory kernel to a special class of analytic functions,
typically exponential and/or exponentially damped cosine functions~\cite{Ceriotti2009a,Ceriotti2010,Morrone2011,Baczewski2013, Stella2014}.
%
% FG: Reference to later results might be helpful to understand this fact
Practically, a fit to oscillatory functions  in time domain can be very difficult since the memory kernels of realistic solute systems usually constitute a mixture of terms with distinct frequencies. 
Importantly the same problem occurs in Laplace domain, see panel b in \Fig{fig:laplace-vs-fourier},
thereby making a fit of the memory kernel directly in Laplace domain not recommendable as well.
In contrast, the signals can be well separated in frequency domain, see panel a therein,
thus simplifying the fit remarkably.
This suggests that a successful method might be formulated in the frequency domain.

\begin{figure}[t]
\includegraphics[width=0.99\columnwidth]{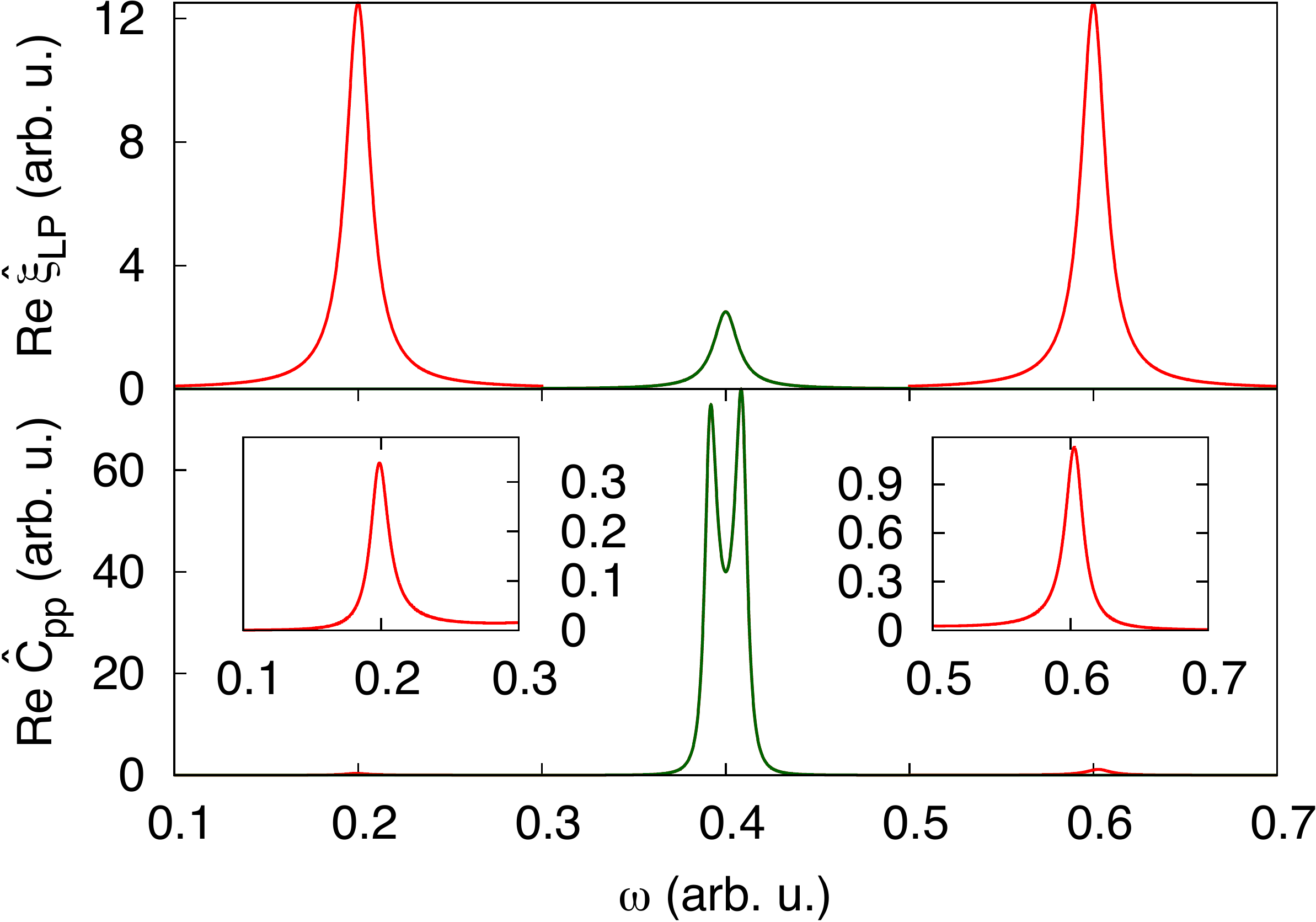}
\caption{\label{fig:resonance}
Two spectral densities, one having only the resonant contribution (green) and the other 
having additionally two off-resonant Lorentzian contributions (red) are depicted in the upper panel.
The corresponding spectra of the harmonic oscillator with frequency $\bar{\omega}=0.4$
coupled to the spectral densities are shown in lower panel.
Insets zoom on the spectral features stemming from the off-resonant contributions in the second spectral density.
}
\end{figure}

Another benefit of working in frequency domain is the possibility to limit the description to the region of the spectral density that is resonant with the system frequency only.
This is illustrated in \Fig{fig:resonance} by comparing the linear absorption spectra of a harmonic oscillator in a bath
described by a spectral density with and without off-resonant contributions.
The two spectra are almost identical and the influence of the off-resonant peaks in the spectral density on spectra
% is not visible without a magnifying glass,
is negligibly small,
 see insets therein, although the off-resonant peak intensity is about five times larger than the intensity in the resonant region. 
This is a manifestation of the well-known fact that the energy can be exchanged efficiently only
when the bath is resonant with the system.
Thus the consideration can be limited to a narrow frequency interval around the system frequency.
In practice, this is a great simplification, since the spectral density might have an extremely elaborate pattern,
if the bath consists of fairly complicated molecules, see, e.g.~panel c) in \Fig{fig:SpectralDensity}.
In time domain it would be impossible to filter out the corresponding signal. 
Importantly it would be equally complicated to extract the frequency window in Laplace domain due to the unsuitable shape of the
Laplace-transformed damped oscillations, see panel b in \Fig{fig:laplace-vs-fourier}.

To this end the problem of parametrizing GLE simulations seems to naturally pose itself in frequency domain,
thereby operating with the spectral density rather than with the memory kernel. 
The transition from Laplace to frequency domain can be easily achieved by setting the Laplace variable imaginary,
i.e.~$s\equiv i\omega$ with a real frequency $\omega$.
Equation~(\ref{eq:Volterra in Laplace domain}) then becomes
\begin{equation}
\label{eq:Volterra in Fourier domain}
i\omega \hat{C}_{pp}(\omega)-C_{pp}(t=0)=-\hat{K}_\mathrm{LP}(\omega)\hat{C}_{pp}(\omega)
\enspace .
\end{equation}
Solving \Eq{eq:Volterra in Fourier domain} for the spectral density yields
\begin{equation}
\label{eq:GetSD}
\hat{K}_\mathrm{LP}(\omega)=\frac{1}{\hat{C}_{pp}(\omega)}-i\omega \enspace ,
\end{equation}
using that $C_{pp}(0)=1$ by normalization;
note that the convolution theorem still applies to the half-sided Fourier transform.

Equation~(\ref{eq:GetSD}) is the starting point of the proposed Fourier-based GLE parametrization scheme,
later referred to as the Fourier method,
that is presented in detail in the next section.

\subsection{Implementation of a Fourier Method}

In this section we elaborate on setting up a  GLE simulation according to \Eq{eq:GLE}
using the previously derived result in \Eq{eq:GetSD} as a starting point.
Two quantities need to be calculated: the effective harmonic frequency, $\bar{\omega}$, defined in \Eq{eq:frequency}
and a spectral density $\xilph(\omega)$.
Since the two memory kernels $\xilp(t)$ and $K_\mathrm{LP}(t)$ differ only by the constant $\bar{\omega}^2$,
the corresponding half-sided Fourier transforms are connected via
\begin{equation}
\label{eq:relation SD}
\hat{K}_\mathrm{LP}(\omega)=\xilph(\omega) + \pi \bar{\omega}^2\delta(\omega) - i\frac{\bar{\omega}^2}{\omega} 
\enspace .
\end{equation}
One notices that the real parts of $\xilph(\omega)$ and $\hat{K}_{\mathrm{LP}}(\omega)$ coincide for $\omega>0$.
Since the memory kernel $\xilp(t)$ is a real function, the knowledge of the real part of $\xilph(\omega)$ and, thus, of 
$\hat{K}_\mathrm{LP}(\omega)$ is sufficient,
because the imaginary part is provided by the Kramer-Kronig relations.
Finally, fitting the hyperbola in the imaginary part of $\hat{K}_\mathrm{LP}(\omega)$ to the fit function $h(\omega)=-d/\omega$
%SI: I inlined the equation.
%\begin{equation}
%\label{eq:fit frequency}
%h(\omega)=-d/\omega
%\end{equation}
provides an easy way to determine the effective harmonic frequency as $\bar{\omega}=\sqrt d$.

In order to use the obtained spectral density in GLE simulations, its corresponding memory kernel $\xilp(t)$
must be given as a superposition of damped cosine functions, $f(t)=2a^2 e^{-bt}\cos(ct),\ \forall a,b,c \in\mathbf{R}, b>0$.
The real part of its Fourier transform reads as a superposition of
\begin{equation}
\label{eq:fit function}
\Re \hat{f}(\omega)=a^2b\cdot \left [ \frac{1}{b^2 + (c-\omega)^2} + \frac{1}{b^2 + (c+\omega)^2} \right ]
\enspace ,
\end{equation}
which can in turn be used to fit the spectral density, $\xilph(\omega)$.
For fitting the hyperbola in the imaginary part, $\Im \hat{K}_\mathrm{LP}(\omega)$, it is recommendable to subtract from it
the imaginary parts of the fit functions given by
% calculated according to Kramer-Kronig relations
%
\begin{equation}
\label{eq:imag fit function}
\Im \hat{f}(\omega)=a^2 \cdot \left[ \frac{c+\omega}{b^2 + (c+\omega)^2} + \frac{c-\omega}{b^2 + (c-\omega)^2} \right]
\enspace .
\end{equation}

\begin{figure}[t]
\includegraphics[width=0.99\columnwidth]{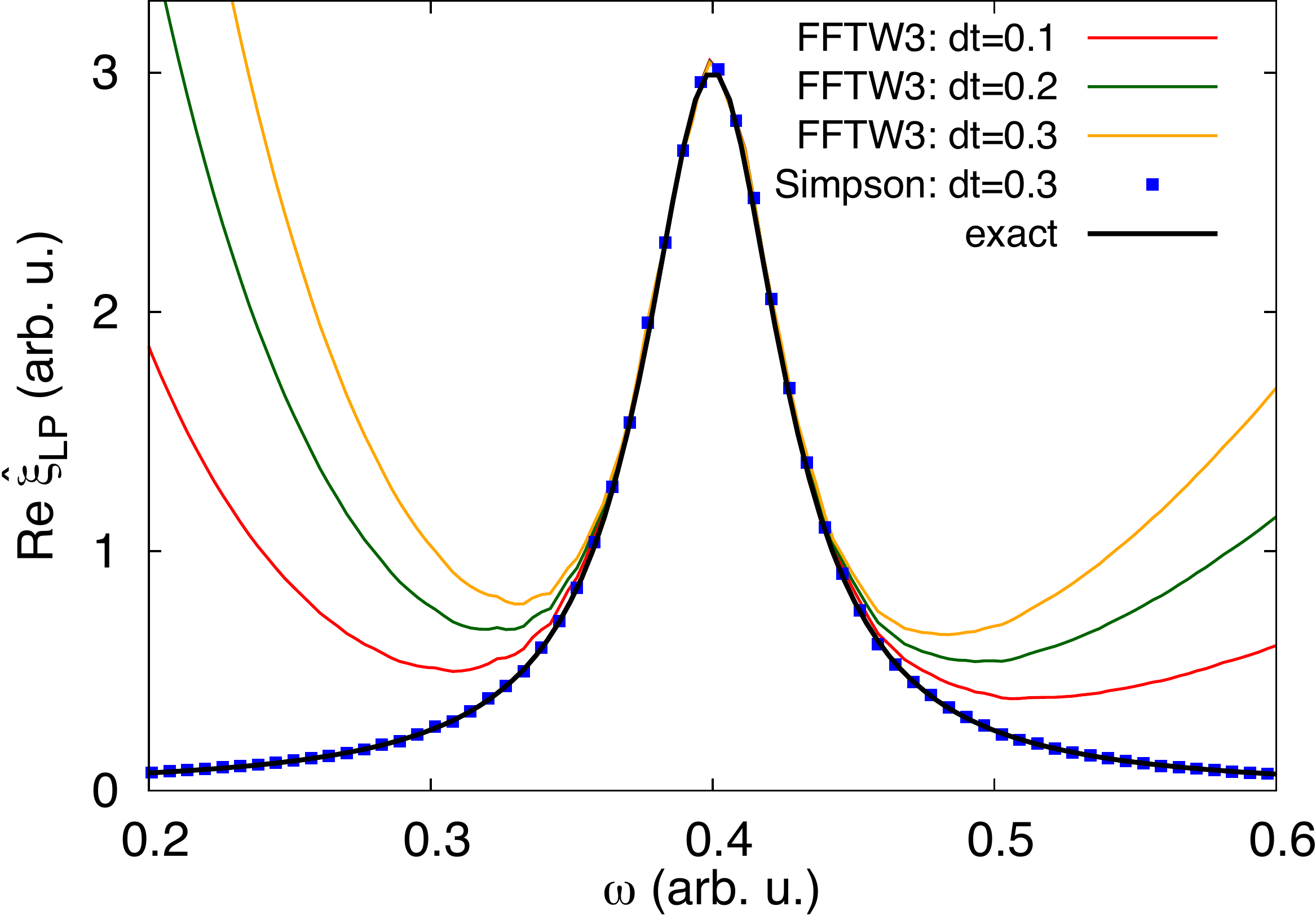}
\caption{\label{fig:FFTW3 errors}
Real part of spectral densities obtained according to \Eq{eq:GetSD} for the test system, see text.
The exact spectral density is shown in black. 
Numerical results obtained using the FFTW3 library for three time steps $0.1, 0.2, 0.3$ are shown in red, green and yellow, respectively.
The result employing the Fourier transform based on the Simpson rule for a time step of $0.3$ is shown as blue dots.}
\end{figure}
\begin{figure}[t]
\includegraphics[width=0.99\columnwidth]{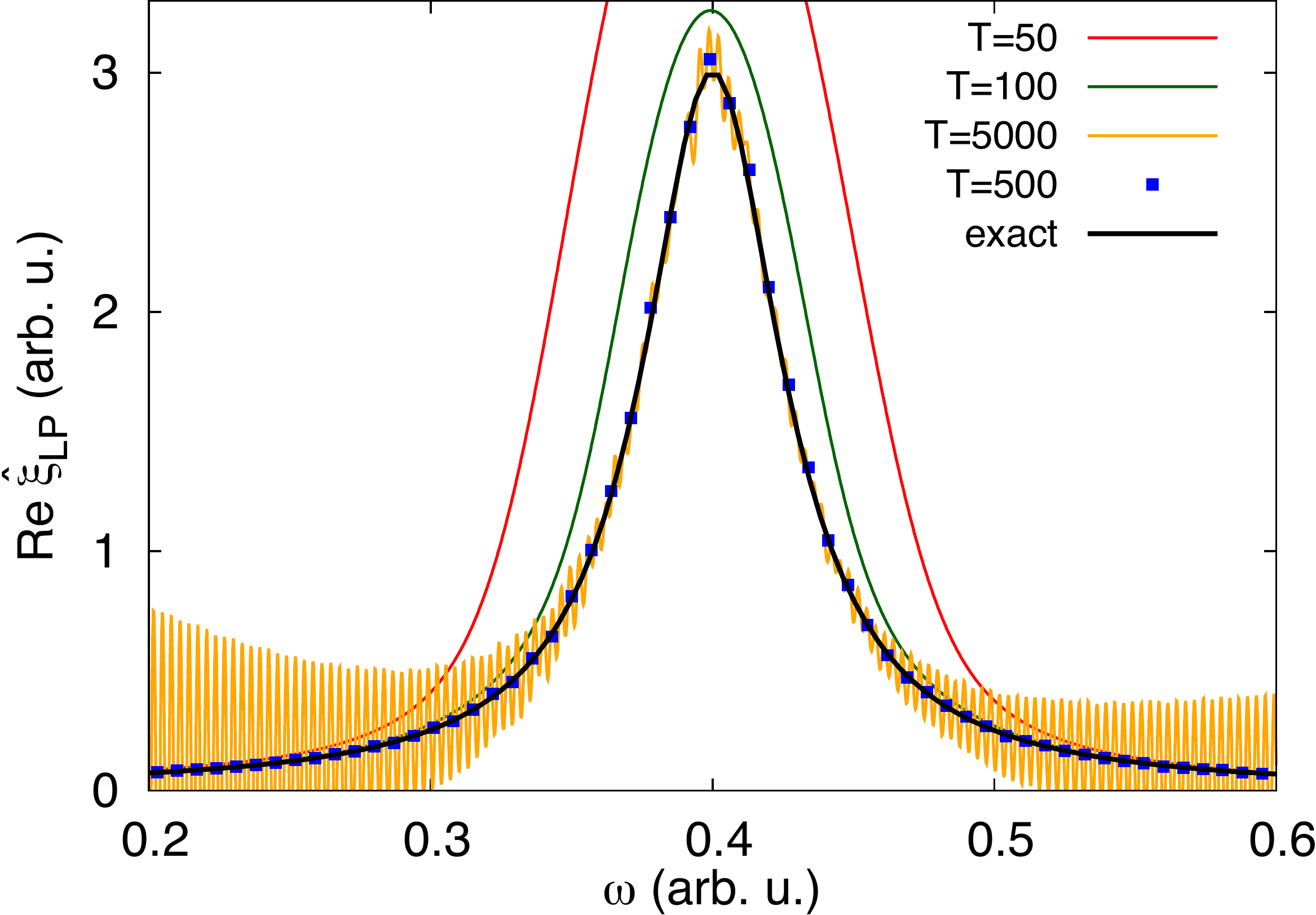}
\caption{\label{fig:Smoothing}
Real part of spectral densities obtained for the test system, see text, upon smoothing the MAF.
The exact spectral density is shown in black. Results for different window widths $T=50, 100, 500, 5000$ are shown as red, green, blue dots and yellow curves, respectively. 
}
%The correlation time of the system is 500.}
\end{figure}

The performance of the method is illustrated on a simple test system: a harmonic oscillator ($\bar{\omega}=0.4$)
in a bath described by a spectral density comprised of a single function as given in \Eq{eq:fit function} with $a=0.03$,
$b=0.03$ and $c=0.4$.
In order to test the procedure, the aforementioned spectral density was used in GLE simulations to produce the MAF, 
that in turn was used to parametrize the spectral density
% according to \Eq{eq:GetSD}
as it was described above.
The resulting spectral density was compared against the input one.

It turns out that the successful use of the procedure presented above rests upon two important numerical issues.
First, the Fourier transform $\hat{C}_{pp}(\omega)$ needs to be calculated very accurately. 
In particular, performing the Fourier transform by means of the standard FFTW3 library~\cite{FFTW05} 
(based on a single-sided sum rule),
led to diverging spectral densities even for moderate timestep sizes ($0.1 - 0.3$), that are typical
for molecular systems with hydrogen atoms, if reasonable units (fs) are employed, see \Fig{fig:FFTW3 errors}.
Here, a more accurate quadrature, such as the 3/8 Simpson integration scheme, yielded excellent results already for a time step of $0.3$.

The second numerical issue concerns the usually insufficiently converged tail of the MAF,
that causes noisy results upon the Fourier transform.
In order to reduce the noise level we suggest Gaussian filtering, that is to multiply the MAF by a Gaussian window function
\begin{equation}
\label{eq:Gaussian window}
G(t)=\exp \left [ -\frac{t^2}{2T^2} \right ] \enspace .
\end{equation}
In frequency domain this corresponds to a convolution with a Gaussian function of the width $\Delta \omega =1/T$
thereby suppressing the noise.
The parameter $T$ should be chosen as a compromise between noise reduction and smoothing errors, see \Fig{fig:Smoothing}.
As a rule of thumb, $T$ can be set equal to the correlation time in the system, which has been $T=500$ for our test system.

%FG: For those who are lazy to read everything
The developed parametrization method  can be summarized by the following steps
\begin{itemize}
  \setlength{\itemsep}{0.05cm}%
  \setlength{\parskip}{0cm}%
%FG: Where/Whether to mention this plan? Here or as a overall result?
\item{Calculate the MAF from explicit MD simulations with sufficient convergence (system dependent)}

\item{Perform a reasonable Gaussian filtering to reduce the noise level upon the Fourier transform,
 taking the correlation time as a starting guess for the window width}
 
\item{Transform MAF into frequency domain using a sufficiently accurate integrator, e.g.~Simpson 3/8 rule}

\item{Calculate the spectral density $\hat{K}_{\mathrm{LP}}(\omega)$ using \Eq{eq:GetSD}}

\item{Estimate $\bar{\omega}$ and fit the real part of $\hat{K}_\mathrm{LP}(\omega)$
in its vicinity to superpositions of functions given in \Eq{eq:fit function}}

\item{Subtract the imaginary counterparts, \Eq{eq:imag fit function} of the fit functions from $\Im \hat{K}_\mathrm{LP}(\omega)$}

\item{Obtain the effective harmonic frequency $\bar{\omega}$ via a hyperbolic fit from the imaginary part of $\hat{K}_\mathrm{LP}(\omega)$}
\end{itemize}

As a word of caution it should be stressed that the presented scheme does not give a correct estimate of a spectral density
at low frequencies due to the divergence of $\hat{K}_\mathrm{LP}(\omega)$ at $\omega=0$ caused by the $\delta$-function,
see \Eq{eq:relation SD}.
This leaves description of pure dephasing times outside reach.

%FG: I don't like having the computational details separated from the Application section. When reading this section without referring to the applications it reads strangely...
\section{Models and computational Details}
\label{CompDet}

We have investigated  two hydrogen-bonded systems -- HOD in H$_2$O, 
and the ionic liquid $[\mathrm{C}_{2}\mbox{mim}][\mbox{NTf}_{2}]$  at room temperature,
as well as an A$_2$ in A model system employed by Berne et al.~\cite{Berne1990}.
These examples have been chosen as water is perhaps the most important solvent and 
HOD features a spectrum with three distinct peaks, which makes it highly suitable for methodological investigations.~\cite{Cho-CR-2008,Kuehn-PRP-2006,Habershon-JCP-2009,CMD-RPMD2,stenger01_027401} 
The ionic liquid represents a system, where, in addition to moderate H-bonding, there exists a strong Coulomb interaction between the ion pairs~\cite{roth12_105026}.
Finally, investigating the A$_2$ in A model system allows us to compare our results with that of Berne et al,
see Ref.~\cite{Berne1990} for the parameters.
%FG: The size of the system was in fact larger, with the same density.

The aqueous systems have been comprised of 466 molecules in a periodic box with the length of $2.4\,$nm
interacting according to the force field adopted from Ref.~\cite{Paesani-JCP-2010}.
The harmonic HOD simulations used in \Sec{sec:A2inA} for comparison, have been performed with exactly the same
setup as the anharmonic ones, with the OH stretching potential in the HOD molecule being harmonic with the frequency
obtained from expanding the respective Morse potential up to the second order ($\mu \omega^2=5081\,$kJ$\cdot$mol$^{-1}$\AA{}$^{-2}$).
The ionic liquid  simulations have been carried out in a periodic box with the length of $4.5\,$nm containing 216 ionic pairs and the force field described in Ref.~\cite{Koeddermann2007}.
All explicit molecular dynamics (MD) simulations have been performed with the GROMACS program package (Version~4.6.5)~\cite{GROMACS}.
The spectra have been computed for the OH stretch in the HOD molecule and for  the C$(2)-$H stretch of the imidazolium ring~\cite{Koeddermann2007} in the ionic liquid.
In the latter case the potential for the C-H stretching motion has been re-parametrized to a Morse potential
using DFT-B3LYP calculations~\cite{Tobias-Master}.
Note that in the spirit of the system-bath treatment the O-H bondlengths have been  taken as the  respective system coordinates, $x$, whereas all other degrees of freedom constituted the bath.
We employed the  ``standard protocol'' for calculating IR spectra, that is, a set of
NVE trajectories, each $6\,$ps long (time step $0.1\,$fs), has been started from
uncorrelated initial conditions sampled from an NVT ensemble.
The dipole autocorrelation functions for the system coordinate have been Fourier-transformed to yield the spectra~\cite{Ivanov-PCCP-2013}.
In order to achieve convergence, 1000 trajectories for the considered stretching motion  have been employed.

For GLE simulations we adopted the method of Colored Noise thermostats~\cite{Ceriotti2009a,Ceriotti2010,Morrone2011},
fitting the spectral density to a superposition of 8-12 functions as given in \Eq{eq:fit function}.
In the GLE simulations all the numerical parameters for the time step, length, number of trajectories,
etc.\ have been the same as in the explicit MD simulations.
Bonds in the rigid bond approach have been fixed using the Settle algorithm as implemented in GROMACS.

\section{Numerical tests}

In this section we first demonstrate that the proposed Fourier method provides
accurate spectral densities that can describe realistic solute dynamics.
Then the performance of the time domain techniques is discussed, using the spectral densities
obtained via the Fourier method as a reference.
Finally, an exceptional case for which the rigid bond approach works,
the A$_2$ in A model system, is discussed in detail.

\subsection{The Fourier method}

In order to test the Fourier method, first the explicit MD simulations are performed
and vibrational spectra and MAFs are calculated.
Then the spectral densities are parametrized from these MAFs according to the Fourier method.
These spectral densities are used as input for implicit GLE simulations, which in turn yield vibrational spectra
that are compared against their explicit MD counterparts.
The coincidence of the GLE and MD spectra manifests the success of the Fourier method.

\begin{figure}[t]
\includegraphics[width=0.99\columnwidth]{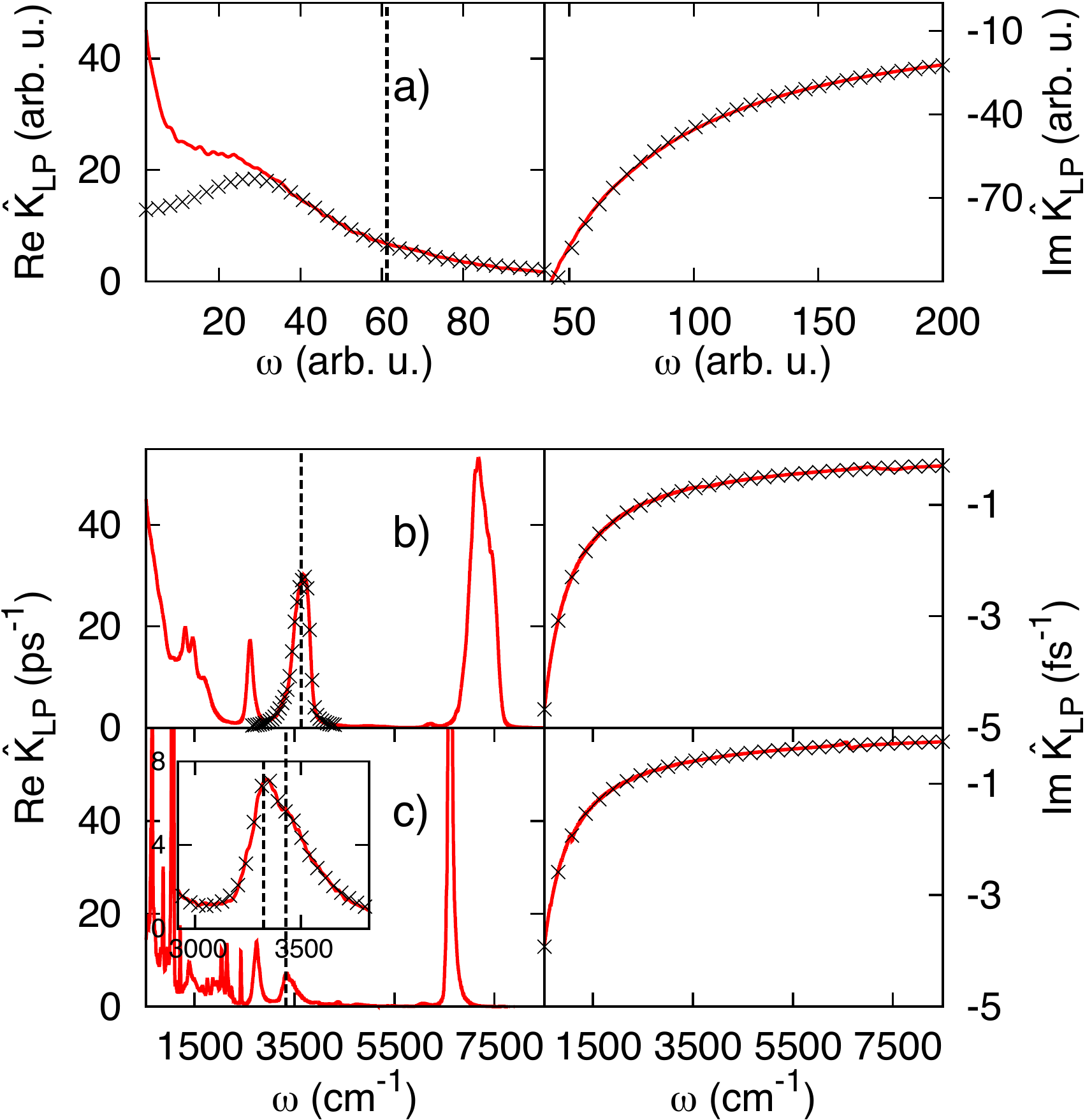}
\caption{\label{fig:SpectralDensity}
Real (left panels) and imaginary (right panels) parts of spectral densities (red curves) obtained according to the 
Fourier method are shown for a) A$_2$ in A, b) HOD in bulk water, and c) the ionic liquid.
Vertical dashed lines denote the effective harmonic frequency $\bar{\omega}$.
The fits to Lorentzian functions, \Eq{eq:fit function}, performed in the resonant region and the
hyperbolic fits to the imaginary parts are depicted with black crosses.
The inset in panel c) zooms into the resonant region.
Note the different scales for imaginary and real parts.
}
\end{figure}

In \Fig{fig:SpectralDensity} the spectral densities, $\hat{K}_\mathrm{LP}(\omega)$, calculated according to \Eq{eq:GetSD},
are shown for the three systems studied.
The real parts of $\hat{K}_\mathrm{LP}(\omega)$ are depicted in the left panels therein.
For the A$_2$ in A model system, panel a), the spectral density nicely reproduces that obtained by Berne et al.~\cite{Berne1990}
The spectral densities for the real systems considered, panels b) and c),
possess peaked contributions at various frequencies
stemming from the coupling to a plethora of vibrational modes in the bath. 
The spectral density is most structured in the ionic liquid case, panel c),
due to the complexity of the molecules involved.
\begin{figure}[t]
\includegraphics[width=0.99\columnwidth]{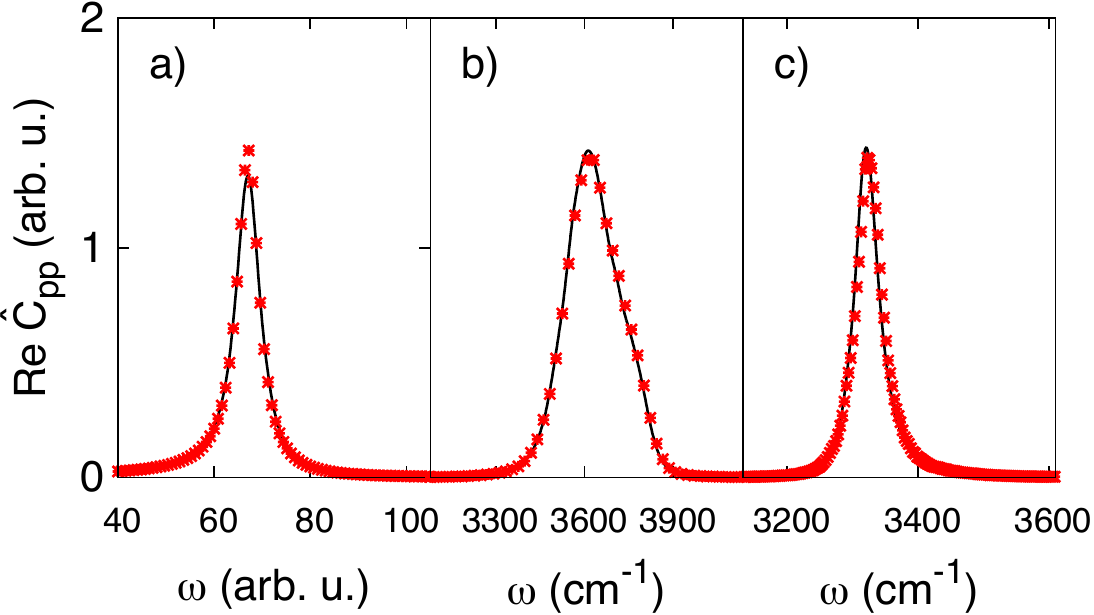}
\caption{\label{fig:Spectra}
Explicit MD (black curve) and GLE spectra (red stars) shown for a) A$_2$ in A, b) HOD in bulk water and c) the ionic liquid.}
\end{figure}
However, as it was explained above, the only important region is in the vicinity of the system frequency $\bar{\omega}$, denoted by a dashed vertical line therein.
The fits to the real part of spectral density in the important region, black crosses in \Fig{fig:SpectralDensity}, 
illustrate the excellent fit quality.
Finally, one sees that the hyperbola in the imaginary part (right panels therein), \Eq{eq:relation SD},
is clearly pronounced and can be thus very well fitted.

Having established the sufficient fit quality for the spectral densities, the GLE simulations are performed
and the resulting vibrational spectra of the three investigated systems (red stars) 
are compared with that obtained from explicit MD simulations (black curves) in \Fig{fig:Spectra}.
It becomes apparent that the GLE simulations excellently reproduce the explicit MD spectra for all systems studied
thereby confirming the success of the Fourier method as a parametrization scheme.

\subsection{Comparison to Time Domain Techniques}
\label{sec:timedomain}
\begin{figure}[t]
\includegraphics[width=0.99\columnwidth]{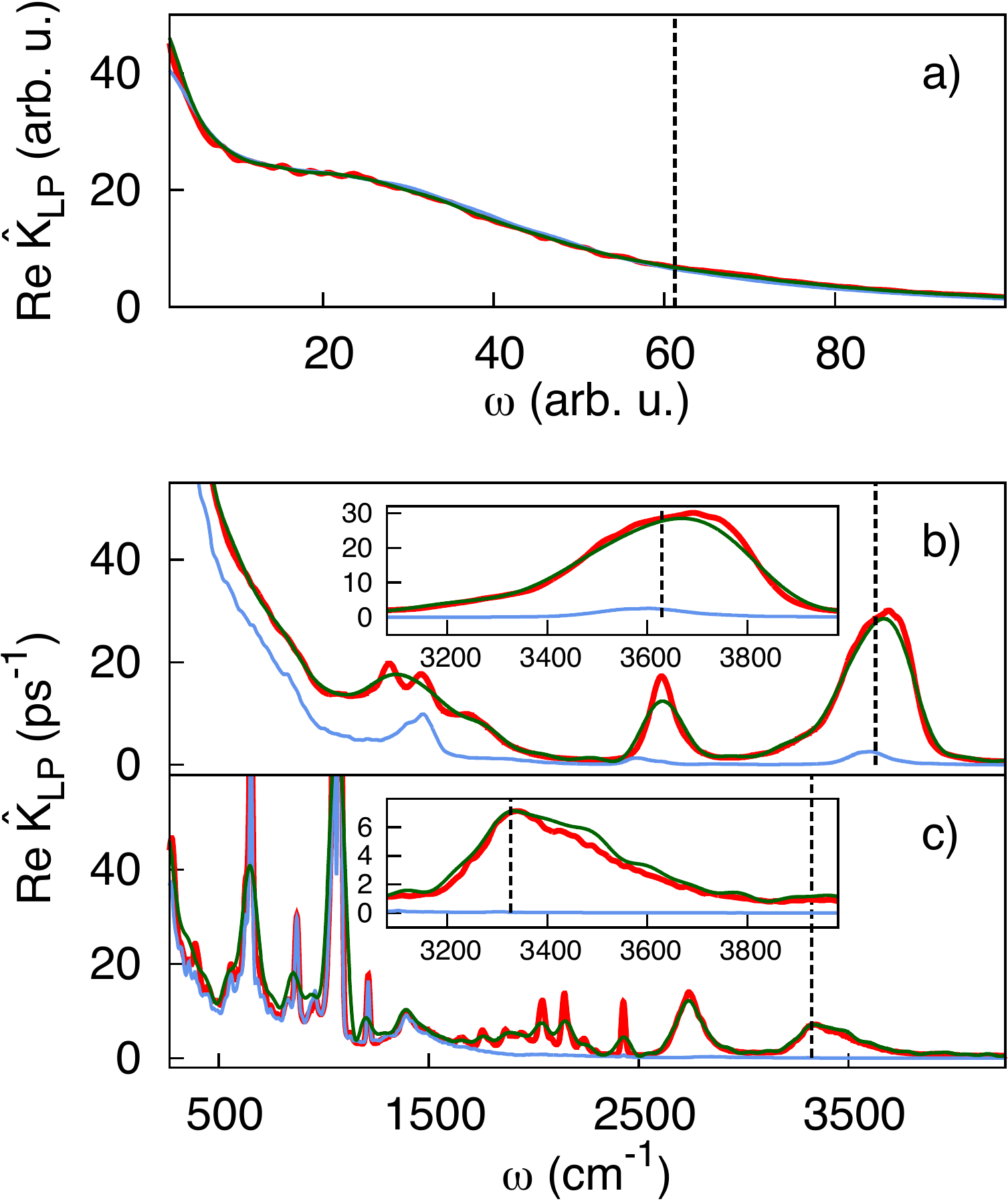}
\caption{\label{fig:spectral densities compared}
Real parts of the spectral densities obtained via the Fourier (red), rigid bond (blue) and Berne and Harp (green) methods
are shown for a) A$_2$ in A b) HOD in bulk water and c) the ionic liquid.
Dashed vertical lines mark $\bar{\omega}$ position.
Insets zoom on the resonant regions.}
\end{figure}

Having obtained the numerical evidence that the spectral densities calculated from the Fourier method
(red lines in \Fig{fig:spectral densities compared}) are reliable, we compare the results obtained via the
rigid bond (blue) and Berne and Harp methods (green) against them.
%
%Berne and Harp
One sees that the real parts of the spectral densities due to the Berne and Harp approach
reproduce the reference results reasonably well.
However, we encountered several numerical issues when using the method for the realistic systems studied.
First, the polynomial interpolation turned out to be very sensitive to the degree $2n$ of the polynomial.
In particular, using $n=1$ was insufficient, $n>2$ led to noticeable spikes causing divergences in the kernel
and only $n=2$ yielded satisfactory results.
Importantly, small inaccuracies in the polynomial interpolation turn out to accumulate strongly
due to repeated integration according to \Eq{eq:diff_Volterraequation}.
Second, even for $n=2$ the kernel diverged at long times and one had to cut it manually at the plateau,
which could not be determined uniquely (data not shown).
These deficiencies together with the general problems of time domain methods discussed in \Sec{sec:Motivation}
make the use of Berne and Harp method not convenient.

%Rigid bond
Considering the results of the rigid bond approach (blue lines in \Fig{fig:spectral densities compared}),
one sees that it is successful only for the A$_2$ in A system, panel a) therein.
The curves corresponding to realistic systems deviate from the reference results significantly.
The general trend is that the larger is the frequency, the more the real part of the spectral density is downscaled.
For HOD in water, panel b), noticeable deviations start from $\approx 1000\,$\cm, implying wrong description
in the region where the bending and OH stretching modes reside.
Remarkably, in the case of the ionic liquid, panel c), the frequency region below 
$1500\,$\cm\ is well reproduced by the rigid system approach.
In contrast, contributions to the important resonant region are completely absent in both cases,
see insets in \Fig{fig:spectral densities compared} for zoom on. 
Since the spectral densities obtained from the Fourier method are proven to be trustworthy,
we thus conclude that the rigid system approach breaks down for the realistic cases considered here.
According to the Laplace domain based analysis by Berne at al.~\cite{Berne1990}, see also \Sec{sec:Rigid bond},
this breakdown is no surprise since the rigid approach is valid only if the system frequency is high compared to that of bath modes.
However, as it was discussed in \Sec{sec:Motivation},
it is more natural to consider the memory kernel in frequency domain.
Here, the formula equivalent to \Eq{eq:RS-LP} reads
\begin{equation}
\xirsh(\omega)=\frac{\xilph(\omega)}{1+i\omega/(\bar{\omega}^2-\omega^2)\xilph(\omega)}
\enspace .
\end{equation}
From this equation it becomes clear that a problem emerges for $\omega \rightarrow \bar{\omega}$,
since the denominator diverges 
under the assumption that  $\xilph(\omega)$ is finite in the vicinity of $\bar{\omega}$;
note this assumption is not fulfilled the irrelevant case of isolated unperturbed systems.
This divergence thus leads to vanishing $\xirsh(\omega)$ in the mostly important resonant region.
Since  $\xirbh(\omega) \approx \xirsh(\omega)$ for sufficiently high frequency $\bar{\omega}$, see \Sec{sec:Rigid bond},
one can conclude that $\xirbh(\omega)$ would vanish in the resonant region as well.
Note that if the latter limit is not yet reached, then one observes a contribution at resonance, which corresponds to the
error $\xirsh(\bar{\omega})-\xirbh(\bar{\omega})$.
This explains our numerical results for the realistic systems studied, namely the absence of the signal for the ionic liquid
and the small but finite contribution to the resonant region observed for the HOD in water case, see insets in panels b) and c). 
Interestingly, this line of reasoning suggests that having the frequency high can only make the situation worse,
as then the error would become smaller and the spectral density even more underestimated.
In fact, considering HOD in D$_2$O, where a frequency separation between the OH stretch and the bath
modes was provided, fully confirmed this conjecture, did not improve the result (data not shown).
Still, the low frequency region, $\omega \ll \bar{\omega}$, comes out correctly as it follows 
from this discussion and has been demonstrated numerically above.

Strictly speaking, the obtained spectral density is generally valid only for the particular (high frequency) mode studied,
thus making this low frequency region irrelevant and the whole rigid bond approach inappropriate.
An important consequence of the generally dramatic underestimation of the spectral density in the resonant region is
the correspondingly overestimated energy relaxation time given in the framework of the Landau-Teller theory~\cite{Tuckerman1993a}
simply as $T_1=1/\xilph(\bar{\omega})$.

One might ask at this point, why the rigid bond method had success for the A$_2$ in A system,
which we elaborate on in the next section.

\subsection{Unexpected success of rigid bond method for A$_2$ in A}
\label{sec:A2inA}

Let us consider what would happen if the system studied was indeed of the CL form.
Then $\xirb(t)=\xid(t)$ without any approximations.
If further the model system potential, $\VSt(x)$, was harmonic then $\xid(t)$ and $\xilp(t)$ would coincide up to a
trivial offset.
Since $\xirb(t)=\xilp(t)$ again up to a constant offset, $\xirs(t)$ needs not to be considered.
It suggests that the success of the rigid bond method for A$_2$ in A is based on the possibility
that this system is indeed of the CL form.

In fact, Berne et al.~\cite{Berne1990} have tested this conjecture by considering the temperature dependence of the
static friction, $\xidh(\omega=0)$ and the frequency renormalisation term, $\xid(t=0)$.
They concluded that the temperature dependence observed excludes the possibility of such a direct  correspondence.
However, the temperature was probed starting just above the melting point, which might have led
to a significant change of the properties.
Additionally, we have recently observed that the failure of such an indirect check
does not necessarily lead to visible discrepancies in the observables, e.g.\ linear vibrational spectra.~\cite{Ivanov-arxiv-2014}
Particularly, testing the linearities of the coupling on both the system and bath sides for the very same realistic systems
suggested that the latter is much more important than the former.
Performing the same test for A$_2$ in A led to a strongly non-linear coupling on the system side and an 
almost ideally linear one on the bath side (data not shown).
Therefore, the important linearity on the bath side suggests that the system can be represented by the CL model.

\begin{figure}[t]
\includegraphics[width=0.99\columnwidth]{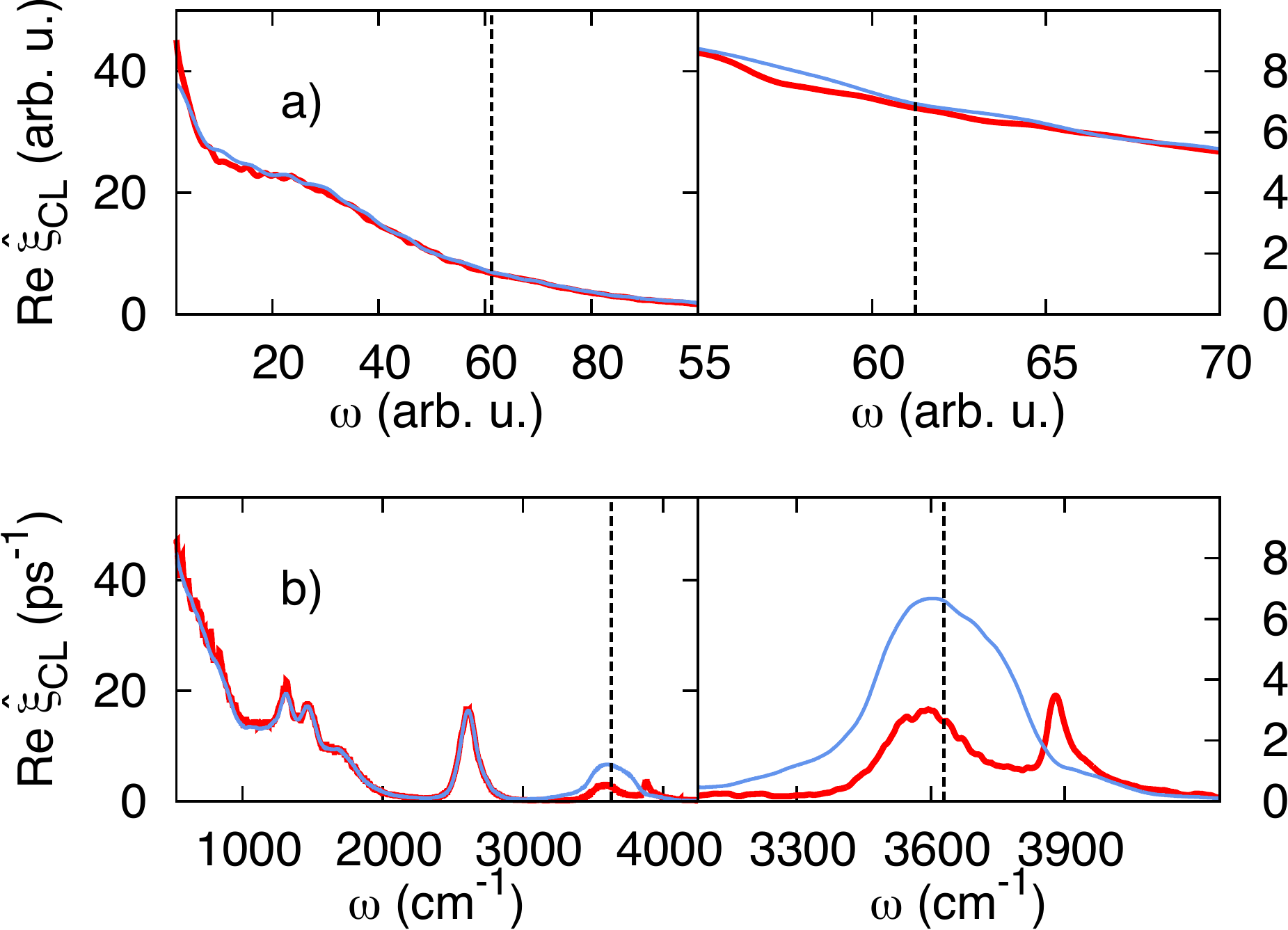}
\caption{\label{fig:CL-check}
Real parts of the spectral densities for harmonic (red) and anharmonic (blue) system potentials
are shown in the left panels for a) A$_2$ in A b) HOD in bulk water.
Right panels zoom on the respective resonant regions.}
\end{figure}

Another possible check is provided by the independence of the spectral density in the CL model from the system potential,
see \Eq{eq:CL-SD}.
This can be tested by computing the memory kernel, $\xid(t)$, for both harmonic and anharmonic system potentials.
It turned out that the spectral densities due to (an)harmonic system potentials coincide for A$_2$ in A,
see panel a) in \Fig{fig:CL-check}.
A reference calculation performed for HOD in water revealed a striking discrepancy in the resonant region, panel b) therein.

These tests considered together indicate that A$_2$ in A can indeed be represented by the CL model.
This also suggests that testing the dependence of the spectral density on the system potential is plausible.

\section{Conclusions and Outlook}

In this paper we considered the problem of parametrizing spectral densities within the linear GLE framework.
We have shown that from the general perspective the problem naturally poses itself in the frequency domain,
and both time and Laplace domains are not intrinsic to the problem.
We have developed a Fourier-based method and compared its performance to the existing time-domain techniques,
that is Berne and Harp~\cite{Berne-ACP-1970} and the widely used rigid bond approach.
It turned out that our method is extremely robust and provides trustworthy results for all systems studied.
In contrast, Berne and Harp approach suffers from numerical problems apart from having general disadvantages of the 
time domain formulations.
Surprisingly the rigid bond method, which is claimed to be accurate in the high frequency limit, turns out to be not
appropriate in general, unless the system studied is of the Caldeira-Leggett form.
Importantly, we have shown that the spectral densities due to the rigid bond method are strongly underestimated in
the resonant region, which makes the common Landau-Teller estimation for the relaxation times questionable.
A parametrizing scheme beyond the linear GLE regime, that is based on \Eq{eq:Volterraequation-CL},
has also been developed and will be published elsewhere.

The authors gratefully acknowledge financial support by the Deutsche Forschungsgemeinschaft (Sfb 652). 

\bibliography{./GLE,./GLE-add}

\begin{thebibliography}{65}%
\makeatletter
\providecommand \@ifxundefined [1]{%
 \@ifx{#1\undefined}
}%
\providecommand \@ifnum [1]{%
 \ifnum #1\expandafter \@firstoftwo
 \else \expandafter \@secondoftwo
 \fi
}%
\providecommand \@ifx [1]{%
 \ifx #1\expandafter \@firstoftwo
 \else \expandafter \@secondoftwo
 \fi
}%
\providecommand \natexlab [1]{#1}%
\providecommand \enquote  [1]{``#1''}%
\providecommand \bibnamefont  [1]{#1}%
\providecommand \bibfnamefont [1]{#1}%
\providecommand \citenamefont [1]{#1}%
\providecommand \href@noop [0]{\@secondoftwo}%
\providecommand \href [0]{\begingroup \@sanitize@url \@href}%
\providecommand \@href[1]{\@@startlink{#1}\@@href}%
\providecommand \@@href[1]{\endgroup#1\@@endlink}%
\providecommand \@sanitize@url [0]{\catcode `\\12\catcode `\$12\catcode
  `\&12\catcode `\#12\catcode `\^12\catcode `\_12\catcode `\%12\relax}%
\providecommand \@@startlink[1]{}%
\providecommand \@@endlink[0]{}%
\providecommand \url  [0]{\begingroup\@sanitize@url \@url }%
\providecommand \@url [1]{\endgroup\@href {#1}{\urlprefix }}%
\providecommand \urlprefix  [0]{URL }%
\providecommand \Eprint [0]{\href }%
\providecommand \doibase [0]{http://dx.doi.org/}%
\providecommand \selectlanguage [0]{\@gobble}%
\providecommand \bibinfo  [0]{\@secondoftwo}%
\providecommand \bibfield  [0]{\@secondoftwo}%
\providecommand \translation [1]{[#1]}%
\providecommand \BibitemOpen [0]{}%
\providecommand \bibitemStop [0]{}%
\providecommand \bibitemNoStop [0]{.\EOS\space}%
\providecommand \EOS [0]{\spacefactor3000\relax}%
\providecommand \BibitemShut  [1]{\csname bibitem#1\endcsname}%
\let\auto@bib@innerbib\@empty
%</preamble>
\bibitem [{\citenamefont {May}\ and\ \citenamefont
  {K\"{u}hn}(2011)}]{Kuehn2011}%
  \BibitemOpen
  \bibfield  {author} {\bibinfo {author} {\bibfnamefont {V.}~\bibnamefont
  {May}}\ and\ \bibinfo {author} {\bibfnamefont {O.}~\bibnamefont {K\"{u}hn}},\
  }\href@noop {} {\emph {\bibinfo {title} {{Charge and Energy Transfer Dynamics
  in Molecular Systems}}}}\ (\bibinfo  {publisher} {Wiley-VCH},\ \bibinfo
  {year} {2011})\BibitemShut {NoStop}%
\bibitem [{\citenamefont {Lemons}(1997)}]{Lemons1997}%
  \BibitemOpen
  \bibfield  {author} {\bibinfo {author} {\bibfnamefont {D.~S.}\ \bibnamefont
  {Lemons}},\ }\href {\doibase 10.1119/1.18725} {\bibfield  {journal} {\bibinfo
   {journal} {Am. J. Phys.}\ }\textbf {\bibinfo {volume} {65}},\ \bibinfo
  {pages} {1079} (\bibinfo {year} {1997})}\BibitemShut {NoStop}%
\bibitem [{\citenamefont {Ornstein}\ and\ \citenamefont
  {Uhlenbeck}(1930)}]{Ornstein1930}%
  \BibitemOpen
  \bibfield  {author} {\bibinfo {author} {\bibfnamefont {L.~S.}\ \bibnamefont
  {Ornstein}}\ and\ \bibinfo {author} {\bibfnamefont {G.~E.}\ \bibnamefont
  {Uhlenbeck}},\ }\href@noop {} {\bibfield  {journal} {\bibinfo  {journal}
  {Phys. Rev.}\ }\textbf {\bibinfo {volume} {36}},\ \bibinfo {pages} {823}
  (\bibinfo {year} {1930})}\BibitemShut {NoStop}%
\bibitem [{\citenamefont {Zwanzig}(2001)}]{ZwanzigBook2001}%
  \BibitemOpen
  \bibfield  {author} {\bibinfo {author} {\bibfnamefont {R.}~\bibnamefont
  {Zwanzig}},\ }\href@noop {} {\emph {\bibinfo {title} {{Nonequilibrium
  Statistical Mechanics}}}}\ (\bibinfo  {publisher} {Oxford University Press},\
  \bibinfo {year} {2001})\BibitemShut {NoStop}%
\bibitem [{\citenamefont {Mori}(1965)}]{Mori1965}%
  \BibitemOpen
  \bibfield  {author} {\bibinfo {author} {\bibfnamefont {H.}~\bibnamefont
  {Mori}},\ }\href@noop {} {\bibfield  {journal} {\bibinfo  {journal} {Prog.
  Theor. Phys.}\ }\textbf {\bibinfo {volume} {33}},\ \bibinfo {pages} {423}
  (\bibinfo {year} {1965})}\BibitemShut {NoStop}%
\bibitem [{\citenamefont {Zwanzig}(1973)}]{Zwanzig1973}%
  \BibitemOpen
  \bibfield  {author} {\bibinfo {author} {\bibfnamefont {R.}~\bibnamefont
  {Zwanzig}},\ }\href {http://link.springer.com/article/10.1007/BF01008729}
  {\bibfield  {journal} {\bibinfo  {journal} {J. Stat. Phys.}\ }\textbf
  {\bibinfo {volume} {9}},\ \bibinfo {pages} {215} (\bibinfo {year}
  {1973})}\BibitemShut {NoStop}%
\bibitem [{\citenamefont {Kawasaki}(1973)}]{Kawasaki1973}%
  \BibitemOpen
  \bibfield  {author} {\bibinfo {author} {\bibfnamefont {K.}~\bibnamefont
  {Kawasaki}},\ }\href {http://iopscience.iop.org/0301-0015/6/9/004} {\bibfield
   {journal} {\bibinfo  {journal} {J. Phys. A Math. Nucl. Gen.}\ }\textbf
  {\bibinfo {volume} {6}},\ \bibinfo {pages} {1289} (\bibinfo {year}
  {1973})}\BibitemShut {NoStop}%
\bibitem [{\citenamefont {Whitnell}\ \emph {et~al.}(1990)\citenamefont
  {Whitnell}, \citenamefont {Wilson},\ and\ \citenamefont
  {Hynes}}]{Whitnell1990}%
  \BibitemOpen
  \bibfield  {author} {\bibinfo {author} {\bibfnamefont {R.~M.}\ \bibnamefont
  {Whitnell}}, \bibinfo {author} {\bibfnamefont {K.~R.}\ \bibnamefont
  {Wilson}}, \ and\ \bibinfo {author} {\bibfnamefont {J.~T.}\ \bibnamefont
  {Hynes}},\ }\href {http://pubs.acs.org/doi/abs/10.1021/j100387a002}
  {\bibfield  {journal} {\bibinfo  {journal} {J. Phys. Chem.}\ }\textbf
  {\bibinfo {volume} {94}},\ \bibinfo {pages} {8625} (\bibinfo {year}
  {1990})}\BibitemShut {NoStop}%
\bibitem [{\citenamefont {Benjamin}\ and\ \citenamefont
  {Whitnell}(1993)}]{Benjamin1993}%
  \BibitemOpen
  \bibfield  {author} {\bibinfo {author} {\bibfnamefont {I.}~\bibnamefont
  {Benjamin}}\ and\ \bibinfo {author} {\bibfnamefont {R.~M.}\ \bibnamefont
  {Whitnell}},\ }\href {\doibase 10.1016/0009-2614(93)85603-L} {\bibfield
  {journal} {\bibinfo  {journal} {Chem. Phys. Lett.}\ }\textbf {\bibinfo
  {volume} {204}},\ \bibinfo {pages} {45} (\bibinfo {year} {1993})}\BibitemShut
  {NoStop}%
\bibitem [{\citenamefont {Tuckerman}\ and\ \citenamefont
  {Berne}(1993)}]{Tuckerman1993a}%
  \BibitemOpen
  \bibfield  {author} {\bibinfo {author} {\bibfnamefont {M.}~\bibnamefont
  {Tuckerman}}\ and\ \bibinfo {author} {\bibfnamefont {B.~J.}\ \bibnamefont
  {Berne}},\ }\href {\doibase 10.1063/1.464723} {\bibfield  {journal} {\bibinfo
   {journal} {J. Chem. Phys.}\ }\textbf {\bibinfo {volume} {98}},\ \bibinfo
  {pages} {7301} (\bibinfo {year} {1993})}\BibitemShut {NoStop}%
\bibitem [{\citenamefont {Gnanakaran}\ and\ \citenamefont
  {Hochstrasser}(1996)}]{Gnanakaran1996a}%
  \BibitemOpen
  \bibfield  {author} {\bibinfo {author} {\bibfnamefont {S.}~\bibnamefont
  {Gnanakaran}}\ and\ \bibinfo {author} {\bibfnamefont {R.~M.}\ \bibnamefont
  {Hochstrasser}},\ }\href {\doibase 10.1063/1.472218} {\bibfield  {journal}
  {\bibinfo  {journal} {J. Chem. Phys.}\ }\textbf {\bibinfo {volume} {105}},\
  \bibinfo {pages} {3486} (\bibinfo {year} {1996})}\BibitemShut {NoStop}%
\bibitem [{\citenamefont {Grote}\ and\ \citenamefont
  {Hynes}(1980)}]{Grote-JCP-1980}%
  \BibitemOpen
  \bibfield  {author} {\bibinfo {author} {\bibfnamefont {R.~F.}\ \bibnamefont
  {Grote}}\ and\ \bibinfo {author} {\bibfnamefont {J.~T.}\ \bibnamefont
  {Hynes}},\ }\href {\doibase 10.1063/1.440485} {\bibfield  {journal} {\bibinfo
   {journal} {J. Chem. Phys.}\ }\textbf {\bibinfo {volume} {73}},\ \bibinfo
  {pages} {2715} (\bibinfo {year} {1980})}\BibitemShut {NoStop}%
\bibitem [{\citenamefont {Ceriotti}\ \emph
  {et~al.}(2009{\natexlab{a}})\citenamefont {Ceriotti}, \citenamefont {Bussi},\
  and\ \citenamefont {Parrinello}}]{Ceriotti2009}%
  \BibitemOpen
  \bibfield  {author} {\bibinfo {author} {\bibfnamefont {M.}~\bibnamefont
  {Ceriotti}}, \bibinfo {author} {\bibfnamefont {G.}~\bibnamefont {Bussi}}, \
  and\ \bibinfo {author} {\bibfnamefont {M.}~\bibnamefont {Parrinello}},\
  }\href {\doibase 10.1103/PhysRevLett.103.030603} {\bibfield  {journal}
  {\bibinfo  {journal} {Phys. Rev. Lett.}\ }\textbf {\bibinfo {volume} {103}},\
  \bibinfo {pages} {030603} (\bibinfo {year} {2009}{\natexlab{a}})}\BibitemShut
  {NoStop}%
\bibitem [{\citenamefont {Ceriotti}\ \emph
  {et~al.}(2009{\natexlab{b}})\citenamefont {Ceriotti}, \citenamefont {Bussi},\
  and\ \citenamefont {Parrinello}}]{Ceriotti2009a}%
  \BibitemOpen
  \bibfield  {author} {\bibinfo {author} {\bibfnamefont {M.}~\bibnamefont
  {Ceriotti}}, \bibinfo {author} {\bibfnamefont {G.}~\bibnamefont {Bussi}}, \
  and\ \bibinfo {author} {\bibfnamefont {M.}~\bibnamefont {Parrinello}},\
  }\href {\doibase 10.1103/PhysRevLett.102.020601} {\bibfield  {journal}
  {\bibinfo  {journal} {Phys. Rev. Lett.}\ }\textbf {\bibinfo {volume} {102}},\
  \bibinfo {pages} {020601} (\bibinfo {year} {2009}{\natexlab{b}})}\BibitemShut
  {NoStop}%
\bibitem [{\citenamefont {Ceriotti}\ \emph {et~al.}(2010)\citenamefont
  {Ceriotti}, \citenamefont {Bussi},\ and\ \citenamefont
  {Parrinello}}]{Ceriotti2010}%
  \BibitemOpen
  \bibfield  {author} {\bibinfo {author} {\bibfnamefont {M.}~\bibnamefont
  {Ceriotti}}, \bibinfo {author} {\bibfnamefont {G.}~\bibnamefont {Bussi}}, \
  and\ \bibinfo {author} {\bibfnamefont {M.}~\bibnamefont {Parrinello}},\
  }\href {\doibase 10.1021/ct900563s} {\bibfield  {journal} {\bibinfo
  {journal} {J. Chem. Theory Comput.}\ }\textbf {\bibinfo {volume} {6}},\
  \bibinfo {pages} {1170} (\bibinfo {year} {2010})}\BibitemShut {NoStop}%
\bibitem [{\citenamefont {Abe}\ \emph {et~al.}(1996)\citenamefont {Abe},
  \citenamefont {Ayik}, \citenamefont {Reinhard},\ and\ \citenamefont
  {Suraud}}]{Abe-PR-1996}%
  \BibitemOpen
  \bibfield  {author} {\bibinfo {author} {\bibfnamefont {Y.}~\bibnamefont
  {Abe}}, \bibinfo {author} {\bibfnamefont {S.}~\bibnamefont {Ayik}}, \bibinfo
  {author} {\bibfnamefont {P.-G.}\ \bibnamefont {Reinhard}}, \ and\ \bibinfo
  {author} {\bibfnamefont {E.}~\bibnamefont {Suraud}},\ }\href {\doibase
  10.1016/0370-1573(96)00003-8} {\bibfield  {journal} {\bibinfo  {journal}
  {Phys. Rep.}\ }\textbf {\bibinfo {volume} {275}},\ \bibinfo {pages} {49}
  (\bibinfo {year} {1996})}\BibitemShut {NoStop}%
\bibitem [{\citenamefont {Tanimura}(2006)}]{Tanimura2006}%
  \BibitemOpen
  \bibfield  {author} {\bibinfo {author} {\bibfnamefont {Y.}~\bibnamefont
  {Tanimura}},\ }\href {\doibase 10.1143/JPSJ.75.082001} {\bibfield  {journal}
  {\bibinfo  {journal} {J. Phys. Soc. Japan}\ }\textbf {\bibinfo {volume}
  {75}},\ \bibinfo {pages} {1} (\bibinfo {year} {2006})}\BibitemShut {NoStop}%
\bibitem [{\citenamefont {Caldeira}\ and\ \citenamefont
  {Leggett}(1981)}]{Caldeira-PRL-1981}%
  \BibitemOpen
  \bibfield  {author} {\bibinfo {author} {\bibfnamefont {A.}~\bibnamefont
  {Caldeira}}\ and\ \bibinfo {author} {\bibfnamefont {A.}~\bibnamefont
  {Leggett}},\ }\href@noop {} {\bibfield  {journal} {\bibinfo  {journal} {Phys.
  Rev. Lett.}\ }\textbf {\bibinfo {volume} {46}},\ \bibinfo {pages} {211}
  (\bibinfo {year} {1981})}\BibitemShut {NoStop}%
\bibitem [{\citenamefont {Caldeira}\ and\ \citenamefont
  {Leggett}(1983{\natexlab{a}})}]{Caldeira-AP-1983}%
  \BibitemOpen
  \bibfield  {author} {\bibinfo {author} {\bibfnamefont {A.~O.}\ \bibnamefont
  {Caldeira}}\ and\ \bibinfo {author} {\bibfnamefont {A.~J.}\ \bibnamefont
  {Leggett}},\ }\href {\doibase 10.1016/0003-4916(83)90202-6} {\bibfield
  {journal} {\bibinfo  {journal} {Ann. Phys. (N. Y).}\ }\textbf {\bibinfo
  {volume} {149}},\ \bibinfo {pages} {374} (\bibinfo {year}
  {1983}{\natexlab{a}})}\BibitemShut {NoStop}%
\bibitem [{\citenamefont {Mukamel}(1995)}]{Mukamel1995}%
  \BibitemOpen
  \bibfield  {author} {\bibinfo {author} {\bibfnamefont {S.}~\bibnamefont
  {Mukamel}},\ }\href@noop {} {\emph {\bibinfo {title} {{Principles of
  Nonlinear Optical Spectroscopy}}}}\ (\bibinfo  {publisher} {Oxford University
  Press, Oxford},\ \bibinfo {year} {1995})\BibitemShut {NoStop}%
\bibitem [{\citenamefont {Palese}\ and\ \citenamefont
  {Mukamel}(1996)}]{Palese1996}%
  \BibitemOpen
  \bibfield  {author} {\bibinfo {author} {\bibfnamefont {S.}~\bibnamefont
  {Palese}}\ and\ \bibinfo {author} {\bibfnamefont {S.}~\bibnamefont
  {Mukamel}},\ }\href {http://pubs.acs.org/doi/abs/10.1021/jp960266l}
  {\bibfield  {journal} {\bibinfo  {journal} {J. Phys. Chem.}\ }\textbf
  {\bibinfo {volume} {3654}},\ \bibinfo {pages} {10380} (\bibinfo {year}
  {1996})}\BibitemShut {NoStop}%
\bibitem [{\citenamefont {Okumura}\ and\ \citenamefont
  {Tanimura}(1997)}]{Okumura1997}%
  \BibitemOpen
  \bibfield  {author} {\bibinfo {author} {\bibfnamefont {K.}~\bibnamefont
  {Okumura}}\ and\ \bibinfo {author} {\bibfnamefont {Y.}~\bibnamefont
  {Tanimura}},\ }\href
  {http://scitation.aip.org/content/aip/journal/jcp/107/7/10.1063/1.474604}
  {\bibfield  {journal} {\bibinfo  {journal} {J. Chem. Phys.}\ }\textbf
  {\bibinfo {volume} {107}},\ \bibinfo {pages} {2267} (\bibinfo {year}
  {1997})}\BibitemShut {NoStop}%
\bibitem [{\citenamefont {Woutersen}\ and\ \citenamefont
  {Bakker}(1999)}]{Woutersen1999}%
  \BibitemOpen
  \bibfield  {author} {\bibinfo {author} {\bibfnamefont {S.}~\bibnamefont
  {Woutersen}}\ and\ \bibinfo {author} {\bibfnamefont {H.~J.}\ \bibnamefont
  {Bakker}},\ }\href {\doibase 10.1103/PhysRevLett.83.2077} {\bibfield
  {journal} {\bibinfo  {journal} {Phys. Rev. Lett.}\ }\textbf {\bibinfo
  {volume} {83}},\ \bibinfo {pages} {2077} (\bibinfo {year}
  {1999})}\BibitemShut {NoStop}%
\bibitem [{\citenamefont {Toutounji}\ and\ \citenamefont
  {Small}(2002)}]{Toutounji2002}%
  \BibitemOpen
  \bibfield  {author} {\bibinfo {author} {\bibfnamefont {M.~M.}\ \bibnamefont
  {Toutounji}}\ and\ \bibinfo {author} {\bibfnamefont {G.~J.}\ \bibnamefont
  {Small}},\ }\href {\doibase 10.1063/1.1495835} {\bibfield  {journal}
  {\bibinfo  {journal} {J. Chem. Phys.}\ }\textbf {\bibinfo {volume} {117}},\
  \bibinfo {pages} {3848} (\bibinfo {year} {2002})}\BibitemShut {NoStop}%
\bibitem [{\citenamefont {Tanimura}\ and\ \citenamefont
  {Ishizaki}(2009)}]{Tanimura2009}%
  \BibitemOpen
  \bibfield  {author} {\bibinfo {author} {\bibfnamefont {Y.}~\bibnamefont
  {Tanimura}}\ and\ \bibinfo {author} {\bibfnamefont {A.}~\bibnamefont
  {Ishizaki}},\ }\href {\doibase 10.1021/ar9000444} {\bibfield  {journal}
  {\bibinfo  {journal} {Acc. Chem. Res.}\ }\textbf {\bibinfo {volume} {42}},\
  \bibinfo {pages} {1270} (\bibinfo {year} {2009})}\BibitemShut {NoStop}%
\bibitem [{\citenamefont {Joutsuka}\ and\ \citenamefont
  {Ando}(2011)}]{Joutsuka2011}%
  \BibitemOpen
  \bibfield  {author} {\bibinfo {author} {\bibfnamefont {T.}~\bibnamefont
  {Joutsuka}}\ and\ \bibinfo {author} {\bibfnamefont {K.}~\bibnamefont
  {Ando}},\ }\href {\doibase 10.1063/1.3594093} {\bibfield  {journal} {\bibinfo
   {journal} {J. Chem. Phys.}\ }\textbf {\bibinfo {volume} {134}},\ \bibinfo
  {pages} {204511} (\bibinfo {year} {2011})}\BibitemShut {NoStop}%
\bibitem [{\citenamefont {Kawai}\ and\ \citenamefont
  {Komatsuzaki}(2011)}]{Kawai2011}%
  \BibitemOpen
  \bibfield  {author} {\bibinfo {author} {\bibfnamefont {S.}~\bibnamefont
  {Kawai}}\ and\ \bibinfo {author} {\bibfnamefont {T.}~\bibnamefont
  {Komatsuzaki}},\ }\href {\doibase 10.1063/1.3561065} {\bibfield  {journal}
  {\bibinfo  {journal} {J. Chem. Phys.}\ }\textbf {\bibinfo {volume} {134}},\
  \bibinfo {pages} {114523} (\bibinfo {year} {2011})}\BibitemShut {NoStop}%
\bibitem [{\citenamefont {Cortes}\ \emph {et~al.}(1985)\citenamefont {Cortes},
  \citenamefont {West},\ and\ \citenamefont {Lindenberg}}]{Cortes1985a}%
  \BibitemOpen
  \bibfield  {author} {\bibinfo {author} {\bibfnamefont {E.}~\bibnamefont
  {Cortes}}, \bibinfo {author} {\bibfnamefont {B.~J.}\ \bibnamefont {West}}, \
  and\ \bibinfo {author} {\bibfnamefont {K.}~\bibnamefont {Lindenberg}},\
  }\href {\doibase 10.1063/1.448268} {\bibfield  {journal} {\bibinfo  {journal}
  {J. Chem. Phys.}\ }\textbf {\bibinfo {volume} {82}},\ \bibinfo {pages} {2708}
  (\bibinfo {year} {1985})}\BibitemShut {NoStop}%
\bibitem [{\citenamefont {Ford}\ and\ \citenamefont
  {Kac}(1987)}]{Ford-JSP-1987}%
  \BibitemOpen
  \bibfield  {author} {\bibinfo {author} {\bibfnamefont {G.~W.}\ \bibnamefont
  {Ford}}\ and\ \bibinfo {author} {\bibfnamefont {M.}~\bibnamefont {Kac}},\
  }\href {\doibase 10.1007/BF01011142} {\bibfield  {journal} {\bibinfo
  {journal} {J. Stat. Phys.}\ }\textbf {\bibinfo {volume} {46}},\ \bibinfo
  {pages} {803} (\bibinfo {year} {1987})}\BibitemShut {NoStop}%
\bibitem [{\citenamefont {Ford}\ \emph {et~al.}(1988)\citenamefont {Ford},
  \citenamefont {Lewis},\ and\ \citenamefont {Oconnell}}]{Ford-PRA-1988}%
  \BibitemOpen
  \bibfield  {author} {\bibinfo {author} {\bibfnamefont {G.~W.}\ \bibnamefont
  {Ford}}, \bibinfo {author} {\bibfnamefont {J.~T.}\ \bibnamefont {Lewis}}, \
  and\ \bibinfo {author} {\bibfnamefont {R.~F.}\ \bibnamefont {Oconnell}},\
  }\href {\doibase 10.1103/PhysRevA.37.4419} {\bibfield  {journal} {\bibinfo
  {journal} {Phys. Rev. A}\ }\textbf {\bibinfo {volume} {37}},\ \bibinfo
  {pages} {4419} (\bibinfo {year} {1988})}\BibitemShut {NoStop}%
\bibitem [{\citenamefont {McDowell}(2000)}]{McDowell2000}%
  \BibitemOpen
  \bibfield  {author} {\bibinfo {author} {\bibfnamefont {H.~K.}\ \bibnamefont
  {McDowell}},\ }\href {\doibase 10.1063/1.481295} {\bibfield  {journal}
  {\bibinfo  {journal} {J. Chem. Phys.}\ }\textbf {\bibinfo {volume} {112}},\
  \bibinfo {pages} {6971} (\bibinfo {year} {2000})}\BibitemShut {NoStop}%
\bibitem [{\citenamefont {Banerjee}\ \emph {et~al.}(2004)\citenamefont
  {Banerjee}, \citenamefont {Bag}, \citenamefont {Banik},\ and\ \citenamefont
  {Ray}}]{Banerjee2004}%
  \BibitemOpen
  \bibfield  {author} {\bibinfo {author} {\bibfnamefont {D.}~\bibnamefont
  {Banerjee}}, \bibinfo {author} {\bibfnamefont {B.~C.}\ \bibnamefont {Bag}},
  \bibinfo {author} {\bibfnamefont {S.~K.}\ \bibnamefont {Banik}}, \ and\
  \bibinfo {author} {\bibfnamefont {D.~S.}\ \bibnamefont {Ray}},\ }\href
  {\doibase 10.1063/1.1711593} {\bibfield  {journal} {\bibinfo  {journal} {J.
  Chem. Phys.}\ }\textbf {\bibinfo {volume} {120}},\ \bibinfo {pages} {8960}
  (\bibinfo {year} {2004})}\BibitemShut {NoStop}%
\bibitem [{\citenamefont {Feynman}\ and\ \citenamefont
  {Vernon}(1963)}]{Feynman-AP-1963}%
  \BibitemOpen
  \bibfield  {author} {\bibinfo {author} {\bibfnamefont {R.}~\bibnamefont
  {Feynman}}\ and\ \bibinfo {author} {\bibfnamefont {F.}~\bibnamefont
  {Vernon}},\ }\href {\doibase 10.1016/0003-4916(63)90068-X} {\bibfield
  {journal} {\bibinfo  {journal} {Ann. Phys. (N. Y).}\ }\textbf {\bibinfo
  {volume} {24}},\ \bibinfo {pages} {118} (\bibinfo {year} {1963})}\BibitemShut
  {NoStop}%
\bibitem [{\citenamefont {Caldeira}\ and\ \citenamefont
  {Leggett}(1983{\natexlab{b}})}]{Caldeira-PhysicaA-1983}%
  \BibitemOpen
  \bibfield  {author} {\bibinfo {author} {\bibfnamefont {A.}~\bibnamefont
  {Caldeira}}\ and\ \bibinfo {author} {\bibfnamefont {A.}~\bibnamefont
  {Leggett}},\ }\href {\doibase 10.1016/0378-4371(83)90013-4} {\bibfield
  {journal} {\bibinfo  {journal} {Phys. A Stat. Mech. its Appl.}\ }\textbf
  {\bibinfo {volume} {121}},\ \bibinfo {pages} {587} (\bibinfo {year}
  {1983}{\natexlab{b}})}\BibitemShut {NoStop}%
\bibitem [{\citenamefont {Dijkstra}\ and\ \citenamefont
  {Tanimura}(2010)}]{Dijkstra-PRL-2010}%
  \BibitemOpen
  \bibfield  {author} {\bibinfo {author} {\bibfnamefont {A.~G.}\ \bibnamefont
  {Dijkstra}}\ and\ \bibinfo {author} {\bibfnamefont {Y.}~\bibnamefont
  {Tanimura}},\ }\href {\doibase 10.1103/PhysRevLett.104.250401} {\bibfield
  {journal} {\bibinfo  {journal} {Phys. Rev. Lett.}\ }\textbf {\bibinfo
  {volume} {104}},\ \bibinfo {pages} {250401} (\bibinfo {year} {2010})},\
  \Eprint {http://arxiv.org/abs/1004.1450} {arXiv:1004.1450} \BibitemShut
  {NoStop}%
\bibitem [{\citenamefont {Suess}\ \emph {et~al.}(2014)\citenamefont {Suess},
  \citenamefont {Eisfeld},\ and\ \citenamefont {Strunz}}]{Suess:2014gz}%
  \BibitemOpen
  \bibfield  {author} {\bibinfo {author} {\bibfnamefont {D.}~\bibnamefont
  {Suess}}, \bibinfo {author} {\bibfnamefont {A.}~\bibnamefont {Eisfeld}}, \
  and\ \bibinfo {author} {\bibfnamefont {W.~T.}\ \bibnamefont {Strunz}},\
  }\href@noop {} {\bibfield  {journal} {\bibinfo  {journal} {Phys. Rev. Lett.}\
  }\textbf {\bibinfo {volume} {113}},\ \bibinfo {pages} {150403} (\bibinfo
  {year} {2014})}\BibitemShut {NoStop}%
\bibitem [{\citenamefont {Giese}\ \emph {et~al.}(2006)\citenamefont {Giese},
  \citenamefont {Petkovi\'{c}}, \citenamefont {Naundorf},\ and\ \citenamefont
  {K\"{u}hn}}]{Kuehn-PRP-2006}%
  \BibitemOpen
  \bibfield  {author} {\bibinfo {author} {\bibfnamefont {K.}~\bibnamefont
  {Giese}}, \bibinfo {author} {\bibfnamefont {M.}~\bibnamefont {Petkovi\'{c}}},
  \bibinfo {author} {\bibfnamefont {H.}~\bibnamefont {Naundorf}}, \ and\
  \bibinfo {author} {\bibfnamefont {O.}~\bibnamefont {K\"{u}hn}},\ }\href
  {\doibase 10.1016/j.physrep.2006.04.005} {\bibfield  {journal} {\bibinfo
  {journal} {Phys. Rep.}\ }\textbf {\bibinfo {volume} {430}},\ \bibinfo {pages}
  {211} (\bibinfo {year} {2006})}\BibitemShut {NoStop}%
\bibitem [{\citenamefont {Egorov}\ \emph {et~al.}(1999)\citenamefont {Egorov},
  \citenamefont {Rabani},\ and\ \citenamefont {Berne}}]{egorov99_5238}%
  \BibitemOpen
  \bibfield  {author} {\bibinfo {author} {\bibfnamefont {S.~A.}\ \bibnamefont
  {Egorov}}, \bibinfo {author} {\bibfnamefont {E.}~\bibnamefont {Rabani}}, \
  and\ \bibinfo {author} {\bibfnamefont {B.~J.}\ \bibnamefont {Berne}},\
  }\href@noop {} {\bibfield  {journal} {\bibinfo  {journal} {J. Chem. Phys.}\
  }\textbf {\bibinfo {volume} {110}},\ \bibinfo {pages} {5238} (\bibinfo {year}
  {1999})}\BibitemShut {NoStop}%
\bibitem [{\citenamefont {Baczewski}\ and\ \citenamefont
  {Bond}(2013)}]{Baczewski2013}%
  \BibitemOpen
  \bibfield  {author} {\bibinfo {author} {\bibfnamefont {A.~D.}\ \bibnamefont
  {Baczewski}}\ and\ \bibinfo {author} {\bibfnamefont {S.~D.}\ \bibnamefont
  {Bond}},\ }\href {\doibase 10.1063/1.4815917} {\bibfield  {journal} {\bibinfo
   {journal} {J. Chem. Phys.}\ }\textbf {\bibinfo {volume} {139}},\ \bibinfo
  {pages} {044107} (\bibinfo {year} {2013})}\BibitemShut {NoStop}%
\bibitem [{\citenamefont {Stella}\ \emph {et~al.}(2014)\citenamefont {Stella},
  \citenamefont {Lorenz},\ and\ \citenamefont {Kantorovich}}]{Stella2014}%
  \BibitemOpen
  \bibfield  {author} {\bibinfo {author} {\bibfnamefont {L.}~\bibnamefont
  {Stella}}, \bibinfo {author} {\bibfnamefont {C.~D.}\ \bibnamefont {Lorenz}},
  \ and\ \bibinfo {author} {\bibfnamefont {L.}~\bibnamefont {Kantorovich}},\
  }\href {\doibase 10.1103/PhysRevB.89.134303} {\bibfield  {journal} {\bibinfo
  {journal} {Phys. Rev. B}\ }\textbf {\bibinfo {volume} {89}},\ \bibinfo
  {pages} {134303} (\bibinfo {year} {2014})}\BibitemShut {NoStop}%
\bibitem [{\citenamefont {Ivanov}\ \emph {et~al.}(2014)\citenamefont {Ivanov},
  \citenamefont {Gottwald},\ and\ \citenamefont {K\"uhn}}]{Ivanov-arxiv-2014}%
  \BibitemOpen
  \bibfield  {author} {\bibinfo {author} {\bibfnamefont {S.~D.}\ \bibnamefont
  {Ivanov}}, \bibinfo {author} {\bibfnamefont {F.}~\bibnamefont {Gottwald}}, \
  and\ \bibinfo {author} {\bibfnamefont {O.}~\bibnamefont {K\"uhn}},\
  }\href@noop {} {\bibfield  {journal} {\bibinfo  {journal} {ArXiv e-prints}\ }
  (\bibinfo {year} {2014})},\ \bibinfo {note} {arXiv:1412.1688
  [physics.chem-ph]},\ \Eprint {http://arxiv.org/abs/0707.3168}
  {arXiv:0707.3168 [hep-th]} \BibitemShut {NoStop}%
\bibitem [{\citenamefont {K\"{u}hn}\ and\ \citenamefont
  {Naundorf}(2003)}]{Kuehn-PCCP-2003}%
  \BibitemOpen
  \bibfield  {author} {\bibinfo {author} {\bibfnamefont {O.}~\bibnamefont
  {K\"{u}hn}}\ and\ \bibinfo {author} {\bibfnamefont {H.}~\bibnamefont
  {Naundorf}},\ }\href {\doibase 10.1039/B209587D} {\bibfield  {journal}
  {\bibinfo  {journal} {Phys. Chem. Chem. Phys.}\ }\textbf {\bibinfo {volume}
  {5}},\ \bibinfo {pages} {79} (\bibinfo {year} {2003})}\BibitemShut {NoStop}%
\bibitem [{\citenamefont {Berne}\ \emph {et~al.}(1990)\citenamefont {Berne},
  \citenamefont {Tuckerman}, \citenamefont {Straub},\ and\ \citenamefont
  {Bug}}]{Berne1990}%
  \BibitemOpen
  \bibfield  {author} {\bibinfo {author} {\bibfnamefont {B.~J.}\ \bibnamefont
  {Berne}}, \bibinfo {author} {\bibfnamefont {M.~E.}\ \bibnamefont
  {Tuckerman}}, \bibinfo {author} {\bibfnamefont {J.~E.}\ \bibnamefont
  {Straub}}, \ and\ \bibinfo {author} {\bibfnamefont {A.~L.~R.}\ \bibnamefont
  {Bug}},\ }\href {\doibase 10.1063/1.458647} {\bibfield  {journal} {\bibinfo
  {journal} {J. Chem. Phys.}\ }\textbf {\bibinfo {volume} {93}},\ \bibinfo
  {pages} {5084} (\bibinfo {year} {1990})}\BibitemShut {NoStop}%
\bibitem [{\citenamefont {Berne}\ and\ \citenamefont
  {Harp}(1970)}]{Berne-ACP-1970}%
  \BibitemOpen
  \bibfield  {author} {\bibinfo {author} {\bibfnamefont {B.~J.}\ \bibnamefont
  {Berne}}\ and\ \bibinfo {author} {\bibfnamefont {G.~D.}\ \bibnamefont
  {Harp}},\ }\href
  {http://books.google.com/books?hl=en\&lr=\&id=vEEO51KNz0IC\&oi=fnd\&pg=PA63\&dq=ON+THE+CALCULATION+OF+TIME+CORRELATION+FUNCTIONS\&ots=jvVmAuw90Y\&sig=z5kmq5GdcXm4IkAOJWhIteiVAag}
  {\bibfield  {journal} {\bibinfo  {journal} {Adv. Chem. Phys}\ }\textbf
  {\bibinfo {volume} {17}},\ \bibinfo {pages} {63} (\bibinfo {year}
  {1970})}\BibitemShut {NoStop}%
\bibitem [{\citenamefont {Berkowitz}\ \emph {et~al.}(1981)\citenamefont
  {Berkowitz}, \citenamefont {Morgan}, \citenamefont {Kouri},\ and\
  \citenamefont {McCammon}}]{Berkowitz1981}%
  \BibitemOpen
  \bibfield  {author} {\bibinfo {author} {\bibfnamefont {M.}~\bibnamefont
  {Berkowitz}}, \bibinfo {author} {\bibfnamefont {J.~D.}\ \bibnamefont
  {Morgan}}, \bibinfo {author} {\bibfnamefont {D.~J.}\ \bibnamefont {Kouri}}, \
  and\ \bibinfo {author} {\bibfnamefont {J.~A.}\ \bibnamefont {McCammon}},\
  }\href {\doibase 10.1063/1.442269} {\bibfield  {journal} {\bibinfo  {journal}
  {J. Chem. Phys.}\ }\textbf {\bibinfo {volume} {75}},\ \bibinfo {pages} {2462}
  (\bibinfo {year} {1981})}\BibitemShut {NoStop}%
\bibitem [{\citenamefont {Lange}\ and\ \citenamefont
  {Grubm\"{u}ller}(2006)}]{Lange2006}%
  \BibitemOpen
  \bibfield  {author} {\bibinfo {author} {\bibfnamefont {O.~F.}\ \bibnamefont
  {Lange}}\ and\ \bibinfo {author} {\bibfnamefont {H.}~\bibnamefont
  {Grubm\"{u}ller}},\ }\href {\doibase 10.1063/1.2199530} {\bibfield  {journal}
  {\bibinfo  {journal} {J. Chem. Phys.}\ }\textbf {\bibinfo {volume} {124}},\
  \bibinfo {pages} {214903} (\bibinfo {year} {2006})}\BibitemShut {NoStop}%
\bibitem [{\citenamefont {Shin}\ \emph {et~al.}(2010)\citenamefont {Shin},
  \citenamefont {Kim}, \citenamefont {Talkner},\ and\ \citenamefont
  {Lee}}]{Shin2010}%
  \BibitemOpen
  \bibfield  {author} {\bibinfo {author} {\bibfnamefont {H.~K.}\ \bibnamefont
  {Shin}}, \bibinfo {author} {\bibfnamefont {C.}~\bibnamefont {Kim}}, \bibinfo
  {author} {\bibfnamefont {P.}~\bibnamefont {Talkner}}, \ and\ \bibinfo
  {author} {\bibfnamefont {E.~K.}\ \bibnamefont {Lee}},\ }\href {\doibase
  10.1016/j.chemphys.2010.05.019} {\bibfield  {journal} {\bibinfo  {journal}
  {Chem. Phys.}\ }\textbf {\bibinfo {volume} {375}},\ \bibinfo {pages} {316}
  (\bibinfo {year} {2010})}\BibitemShut {NoStop}%
\bibitem [{\citenamefont {Straub}\ \emph {et~al.}(1987)\citenamefont {Straub},
  \citenamefont {Borkovec},\ and\ \citenamefont {Berne}}]{Straub1987}%
  \BibitemOpen
  \bibfield  {author} {\bibinfo {author} {\bibfnamefont {J.~E.}\ \bibnamefont
  {Straub}}, \bibinfo {author} {\bibfnamefont {M.}~\bibnamefont {Borkovec}}, \
  and\ \bibinfo {author} {\bibfnamefont {B.~J.}\ \bibnamefont {Berne}},\ }\href
  {http://pubs.acs.org/doi/abs/10.1021/j100303a019} {\bibfield  {journal}
  {\bibinfo  {journal} {J. Phys. Chem.}\ }\textbf {\bibinfo {volume} {91}},\
  \bibinfo {pages} {4995} (\bibinfo {year} {1987})}\BibitemShut {NoStop}%
\bibitem [{\citenamefont {Goodyear}\ and\ \citenamefont
  {Stratt}(1996)}]{Goodyear1996}%
  \BibitemOpen
  \bibfield  {author} {\bibinfo {author} {\bibfnamefont {G.}~\bibnamefont
  {Goodyear}}\ and\ \bibinfo {author} {\bibfnamefont {R.~M.}\ \bibnamefont
  {Stratt}},\ }\href {\doibase 10.1063/1.472835} {\bibfield  {journal}
  {\bibinfo  {journal} {J. Chem. Phys.}\ }\textbf {\bibinfo {volume} {105}},\
  \bibinfo {pages} {10050} (\bibinfo {year} {1996})}\BibitemShut {NoStop}%
\bibitem [{\citenamefont {Berkowitz}\ \emph {et~al.}(1983)\citenamefont
  {Berkowitz}, \citenamefont {Morgan},\ and\ \citenamefont
  {McCammon}}]{Berkowitz1983}%
  \BibitemOpen
  \bibfield  {author} {\bibinfo {author} {\bibfnamefont {M.}~\bibnamefont
  {Berkowitz}}, \bibinfo {author} {\bibfnamefont {J.~D.}\ \bibnamefont
  {Morgan}}, \ and\ \bibinfo {author} {\bibfnamefont {J.~A.}\ \bibnamefont
  {McCammon}},\ }\href {\doibase 10.1063/1.445244} {\bibfield  {journal}
  {\bibinfo  {journal} {J. Chem. Phys.}\ }\textbf {\bibinfo {volume} {78}},\
  \bibinfo {pages} {3256} (\bibinfo {year} {1983})}\BibitemShut {NoStop}%
\bibitem [{\citenamefont {Grabert}\ \emph {et~al.}(1988)\citenamefont
  {Grabert}, \citenamefont {Schramm},\ and\ \citenamefont
  {Ingold}}]{grabert88_115}%
  \BibitemOpen
  \bibfield  {author} {\bibinfo {author} {\bibfnamefont {H.}~\bibnamefont
  {Grabert}}, \bibinfo {author} {\bibfnamefont {P.}~\bibnamefont {Schramm}}, \
  and\ \bibinfo {author} {\bibfnamefont {G.-L.}\ \bibnamefont {Ingold}},\
  }\href@noop {} {\bibfield  {journal} {\bibinfo  {journal} {Phys. Rep.}\
  }\textbf {\bibinfo {volume} {168}},\ \bibinfo {pages} {115} (\bibinfo {year}
  {1988})}\BibitemShut {NoStop}%
\bibitem [{\citenamefont {Todd}(1962)}]{Todd1962}%
  \BibitemOpen
  \bibfield  {author} {\bibinfo {author} {\bibfnamefont {J.}~\bibnamefont
  {Todd}},\ }\href@noop {} {\emph {\bibinfo {title} {{Survey of Numerical
  Analysis}}}}\ (\bibinfo  {publisher} {McGraw-Hill},\ \bibinfo {year}
  {1962})\BibitemShut {NoStop}%
\bibitem [{\citenamefont {Day}(1967)}]{Day1967}%
  \BibitemOpen
  \bibfield  {author} {\bibinfo {author} {\bibfnamefont {J.~T.}\ \bibnamefont
  {Day}},\ }\href@noop {} {\bibfield  {journal} {\bibinfo  {journal} {Bit}\
  }\textbf {\bibinfo {volume} {7}},\ \bibinfo {pages} {71} (\bibinfo {year}
  {1967})}\BibitemShut {NoStop}%
\bibitem [{\citenamefont {Morrone}\ \emph {et~al.}(2011)\citenamefont
  {Morrone}, \citenamefont {Markland}, \citenamefont {Ceriotti},\ and\
  \citenamefont {Berne}}]{Morrone2011}%
  \BibitemOpen
  \bibfield  {author} {\bibinfo {author} {\bibfnamefont {J.~A.}\ \bibnamefont
  {Morrone}}, \bibinfo {author} {\bibfnamefont {T.~E.}\ \bibnamefont
  {Markland}}, \bibinfo {author} {\bibfnamefont {M.}~\bibnamefont {Ceriotti}},
  \ and\ \bibinfo {author} {\bibfnamefont {B.~J.}\ \bibnamefont {Berne}},\
  }\href {\doibase 10.1063/1.3518369} {\bibfield  {journal} {\bibinfo
  {journal} {J. Chem. Phys.}\ }\textbf {\bibinfo {volume} {134}},\ \bibinfo
  {pages} {014103} (\bibinfo {year} {2011})}\BibitemShut {NoStop}%
\bibitem [{\citenamefont {Frigo}\ and\ \citenamefont {Johnson}(2005)}]{FFTW05}%
  \BibitemOpen
  \bibfield  {author} {\bibinfo {author} {\bibfnamefont {M.}~\bibnamefont
  {Frigo}}\ and\ \bibinfo {author} {\bibfnamefont {S.~G.}\ \bibnamefont
  {Johnson}},\ }\href@noop {} {\bibfield  {journal} {\bibinfo  {journal}
  {Proceedings of the IEEE}\ }\textbf {\bibinfo {volume} {93}},\ \bibinfo
  {pages} {216} (\bibinfo {year} {2005})},\ \bibinfo {note} {special issue on
  ``Program Generation, Optimization, and Platform Adaptation''}\BibitemShut
  {NoStop}%
\bibitem [{\citenamefont {Cho}(2008)}]{Cho-CR-2008}%
  \BibitemOpen
  \bibfield  {author} {\bibinfo {author} {\bibfnamefont {M.}~\bibnamefont
  {Cho}},\ }\href {\doibase 10.1021/cr078377b} {\bibfield  {journal} {\bibinfo
  {journal} {Chem. Rev.}\ }\textbf {\bibinfo {volume} {108}},\ \bibinfo {pages}
  {1331} (\bibinfo {year} {2008})}\BibitemShut {NoStop}%
\bibitem [{\citenamefont {Habershon}\ and\ \citenamefont
  {Manolopoulos}(2009)}]{Habershon-JCP-2009}%
  \BibitemOpen
  \bibfield  {author} {\bibinfo {author} {\bibfnamefont {S.}~\bibnamefont
  {Habershon}}\ and\ \bibinfo {author} {\bibfnamefont {D.~E.}\ \bibnamefont
  {Manolopoulos}},\ }\href {\doibase 10.1063/1.3276109} {\bibfield  {journal}
  {\bibinfo  {journal} {J. Chem. Phys.}\ }\textbf {\bibinfo {volume} {131}},\
  \bibinfo {pages} {244518} (\bibinfo {year} {2009})}\BibitemShut {NoStop}%
\bibitem [{\citenamefont {Ivanov}\ \emph {et~al.}(2010)\citenamefont {Ivanov},
  \citenamefont {Witt}, \citenamefont {Shiga},\ and\ \citenamefont
  {Marx}}]{CMD-RPMD2}%
  \BibitemOpen
  \bibfield  {author} {\bibinfo {author} {\bibfnamefont {S.~D.}\ \bibnamefont
  {Ivanov}}, \bibinfo {author} {\bibfnamefont {A.}~\bibnamefont {Witt}},
  \bibinfo {author} {\bibfnamefont {M.}~\bibnamefont {Shiga}}, \ and\ \bibinfo
  {author} {\bibfnamefont {D.}~\bibnamefont {Marx}},\ }\href@noop {} {\bibfield
   {journal} {\bibinfo  {journal} {J.\ Chem.\ Phys.}\ }\textbf {\bibinfo
  {volume} {132}},\ \bibinfo {pages} {031101} (\bibinfo {year}
  {2010})}\BibitemShut {NoStop}%
\bibitem [{\citenamefont {Stenger}\ \emph {et~al.}(2001)\citenamefont
  {Stenger}, \citenamefont {Madsen}, \citenamefont {Hamm}, \citenamefont
  {Nibbering},\ and\ \citenamefont {Elsaesser}}]{stenger01_027401}%
  \BibitemOpen
  \bibfield  {author} {\bibinfo {author} {\bibfnamefont {J.}~\bibnamefont
  {Stenger}}, \bibinfo {author} {\bibfnamefont {D.}~\bibnamefont {Madsen}},
  \bibinfo {author} {\bibfnamefont {P.}~\bibnamefont {Hamm}}, \bibinfo {author}
  {\bibfnamefont {E.~T.~J.}\ \bibnamefont {Nibbering}}, \ and\ \bibinfo
  {author} {\bibfnamefont {T.}~\bibnamefont {Elsaesser}},\ }\href@noop {}
  {\bibfield  {journal} {\bibinfo  {journal} {Phys. Rev. Lett.}\ }\textbf
  {\bibinfo {volume} {87}},\ \bibinfo {pages} {027401} (\bibinfo {year}
  {2001})}\BibitemShut {NoStop}%
\bibitem [{\citenamefont {Roth}\ \emph {et~al.}(2012)\citenamefont {Roth},
  \citenamefont {Chatzipapadopoulos}, \citenamefont {Kerl\'e}, \citenamefont
  {Friedriszik}, \citenamefont {L{\"u}tgens}, \citenamefont {Lochbrunner},
  \citenamefont {K{\"u}hn},\ and\ \citenamefont {Ludwig}}]{roth12_105026}%
  \BibitemOpen
  \bibfield  {author} {\bibinfo {author} {\bibfnamefont {C.}~\bibnamefont
  {Roth}}, \bibinfo {author} {\bibfnamefont {S.}~\bibnamefont
  {Chatzipapadopoulos}}, \bibinfo {author} {\bibfnamefont {D.}~\bibnamefont
  {Kerl\'e}}, \bibinfo {author} {\bibfnamefont {F.}~\bibnamefont
  {Friedriszik}}, \bibinfo {author} {\bibfnamefont {M.}~\bibnamefont
  {L{\"u}tgens}}, \bibinfo {author} {\bibfnamefont {S.}~\bibnamefont
  {Lochbrunner}}, \bibinfo {author} {\bibfnamefont {O.}~\bibnamefont
  {K{\"u}hn}}, \ and\ \bibinfo {author} {\bibfnamefont {R.}~\bibnamefont
  {Ludwig}},\ }\href@noop {} {\bibfield  {journal} {\bibinfo  {journal} {New J.
  Phys.}\ }\textbf {\bibinfo {volume} {14}},\ \bibinfo {pages} {105026}
  (\bibinfo {year} {2012})}\BibitemShut {NoStop}%
\bibitem [{\citenamefont {Paesani}\ and\ \citenamefont
  {Voth}(2010)}]{Paesani-JCP-2010}%
  \BibitemOpen
  \bibfield  {author} {\bibinfo {author} {\bibfnamefont {F.}~\bibnamefont
  {Paesani}}\ and\ \bibinfo {author} {\bibfnamefont {G.~A.}\ \bibnamefont
  {Voth}},\ }\href {\doibase 10.1063/1.3291212} {\bibfield  {journal} {\bibinfo
   {journal} {J. Chem. Phys.}\ }\textbf {\bibinfo {volume} {132}},\ \bibinfo
  {pages} {014105} (\bibinfo {year} {2010})}\BibitemShut {NoStop}%
\bibitem [{\citenamefont {K\"{o}ddermann}\ \emph {et~al.}(2007)\citenamefont
  {K\"{o}ddermann}, \citenamefont {Paschek},\ and\ \citenamefont
  {Ludwig}}]{Koeddermann2007}%
  \BibitemOpen
  \bibfield  {author} {\bibinfo {author} {\bibfnamefont {T.}~\bibnamefont
  {K\"{o}ddermann}}, \bibinfo {author} {\bibfnamefont {D.}~\bibnamefont
  {Paschek}}, \ and\ \bibinfo {author} {\bibfnamefont {R.}~\bibnamefont
  {Ludwig}},\ }\href {\doibase 10.1002/cphc.200700552} {\bibfield  {journal}
  {\bibinfo  {journal} {{ChemPhysChem}}\ }\textbf {\bibinfo {volume} {8}},\
  \bibinfo {pages} {2464} (\bibinfo {year} {2007})}\BibitemShut {NoStop}%
\bibitem [{\citenamefont {Hess}\ \emph {et~al.}(2008)\citenamefont {Hess},
  \citenamefont {Kutzner}, \citenamefont {van~der Spoel},\ and\ \citenamefont
  {Lindahl}}]{GROMACS}%
  \BibitemOpen
  \bibfield  {author} {\bibinfo {author} {\bibfnamefont {B.}~\bibnamefont
  {Hess}}, \bibinfo {author} {\bibfnamefont {C.}~\bibnamefont {Kutzner}},
  \bibinfo {author} {\bibfnamefont {D.}~\bibnamefont {van~der Spoel}}, \ and\
  \bibinfo {author} {\bibfnamefont {E.}~\bibnamefont {Lindahl}},\ }\href
  {\doibase 10.1021/ct700301q} {\bibfield  {journal} {\bibinfo  {journal} {J.
  Chem. Theory Comput.}\ }\textbf {\bibinfo {volume} {4}},\ \bibinfo {pages}
  {435} (\bibinfo {year} {2008})}\BibitemShut {NoStop}%
\bibitem [{\citenamefont {Zentel}(2012)}]{Tobias-Master}%
  \BibitemOpen
  \bibfield  {author} {\bibinfo {author} {\bibfnamefont {T.}~\bibnamefont
  {Zentel}},\ }\emph {\bibinfo {title} {{(Non-)linear Spectroscopy Based on
  Classical Trajectories}}},\ \href@noop {} {Master's thesis},\ \bibinfo
  {school} {Rostock University}, \bibinfo {address} {Rostock, Germany}
  (\bibinfo {year} {2012}),\ \bibinfo {note} {uRL:
  \texttt{http://rosdok.uni-rostock.de/resolve/id/rosdok\_thesis\_0000000013}}\BibitemShut
  {NoStop}%
\bibitem [{\citenamefont {Ivanov}\ \emph {et~al.}(2013)\citenamefont {Ivanov},
  \citenamefont {Witt},\ and\ \citenamefont {Marx}}]{Ivanov-PCCP-2013}%
  \BibitemOpen
  \bibfield  {author} {\bibinfo {author} {\bibfnamefont {S.~D.}\ \bibnamefont
  {Ivanov}}, \bibinfo {author} {\bibfnamefont {A.}~\bibnamefont {Witt}}, \ and\
  \bibinfo {author} {\bibfnamefont {D.}~\bibnamefont {Marx}},\ }\href {\doibase
  10.1039/c3cp44523b} {\bibfield  {journal} {\bibinfo  {journal} {Phys. Chem.
  Chem. Phys.}\ }\textbf {\bibinfo {volume} {15}},\ \bibinfo {pages} {10270}
  (\bibinfo {year} {2013})}\BibitemShut {NoStop}%
\end{thebibliography}%

\end{document}